\documentclass[twocolumn,superscriptaddress,preprintnumbers,amsmath,10pt,aps,prl,nobibnotes,longbibliography]{revtex4-1} 

\usepackage{algorithm, algorithmic}

\usepackage{amsmath}
\usepackage{amsfonts}
\usepackage{bm}
\usepackage{color}
\usepackage{verbatim}
\usepackage{graphicx}
\usepackage{lipsum}
\usepackage[utf8]{inputenc}
\usepackage{times}
\usepackage{graphicx}
\usepackage{bm}
\usepackage{amssymb,multirow}
\usepackage[dvipsnames]{xcolor}
\usepackage{soul}
\usepackage[normalem]{ulem} 
\usepackage[mathlines]{lineno}
\usepackage{bbm}
\usepackage{siunitx}
\usepackage{array}
\usepackage{upgreek}
\usepackage{comment}
\usepackage{hyperref}
\usepackage{commath}
\hypersetup{
    colorlinks=true,
    linkcolor=blue,
    citecolor=blue,
    urlcolor=blue
    }
\usepackage{subcaption}
\usepackage{graphicx}
\usepackage{caption}
\usepackage{balance}

\usepackage{ragged2e} \makeatletter \long\def\@makecaption#1#2{ \par \begingroup \small \parindent=0pt \justifying \noindent #1\ #2\par \endgroup } \makeatother

\def\Vecm{\mathbf{m}}

\def\Veco{\mathbf{o}}
\def\Vecp{\mathbf{p}}

\def\Vecs{\mathbf{s}}
\def\Vect{\mathbf{t}}
\def\Vecu{\mathbf{u}}
\def\Vecv{\mathbf{v}}

\def\VecB{\mathbf{B}}

\def\VecD{\mathbf{D}}

\def\VecI{\mathbf{I}}

\def\VecH{\mathbf{H}}

\def\VecP{\mathbf{P}}

\def\VecR{\mathbf{R}}

\def\VecT{\mathbf{T}}
\def\VecU{\mathbf{U}}
\def\VecV{\mathbf{V}}

\begin{document}

\title{Adaptive wavefront shaping through flexible multicore photonic lantern fibres}

\author{Robert~J.~Kilpatrick}
\thanks{These authors contributed equally to this work}
\affiliation{Department of Physics and Astronomy, University of Exeter, Exeter, EX4 4QL. UK.}
\author{Alasdair Milne}
\thanks{These authors contributed equally to this work}
\affiliation{Institute of Photonics and Quantum Sciences, Heriot-Watt University, Edinburgh, EH14 4AS UK.}
\author{Kerrianne Harrington}
\affiliation{Centre for Photonics and Photonic Materials, Department of Physics, University of Bath, Claverton Down, Bath, BA2 7AY, UK.}
\author{Thomas~A.~Wright}
\affiliation{Centre for Photonics and Photonic Materials, Department of Physics, University of Bath, Claverton Down, Bath, BA2 7AY, UK.}
\author{José~C.~A.~Rocha}
\affiliation{Department of Physics and Astronomy, University of Exeter, Exeter, EX4 4QL. UK.}
\author{Aurélien~Benoît}
\affiliation{Institute of Photonics and Quantum Sciences, Heriot-Watt University, Edinburgh, EH14 4AS UK.}
\author{Helen Parker}
\affiliation{Institute of Photonics and Quantum Sciences, Heriot-Watt University, Edinburgh, EH14 4AS UK.}
\author{Chaitanya~K.~Mididoddi}
\affiliation{Department of Physics and Astronomy, University of Exeter, Exeter, EX4 4QL. UK.}
\author{James~M.~Stone}
\affiliation{Centre for Photonics and Photonic Materials, Department of Physics, University of Bath, Claverton Down, Bath, BA2 7AY, UK.}
\author{Tim~A.~Birks}
\email{pystab@bath.ac.uk}
\affiliation{Centre for Photonics and Photonic Materials, Department of Physics, University of Bath, Claverton Down, Bath, BA2 7AY, UK.}
\author{Robert~R.~Thomson}
\email{R.R.Thomson@hw.ac.uk}
\affiliation{Institute of Photonics and Quantum Sciences, Heriot-Watt University, Edinburgh, EH14 4AS UK.}
\author{David~B.~Phillips}
\email{d.phillips@exeter.ac.uk}
\affiliation{Department of Physics and Astronomy, University of Exeter, Exeter, EX4 4QL. UK.}

\begin{abstract}
Holographic micro-endoscopy enables images to be funnelled all-optically through hair-thin strands of optical fibre. Future applications of this technology encompass biomedical imaging within mobile organisms, LiDAR, and industrial inspection. Yet holographic imaging through {\it dynamically flexible} optical fibres, called for in many real-world scenarios, represents an ongoing challenge. In this work we demonstrate how a bespoke fibre architecture that we term a `photonic lantern fibre' offers a way forward. This device benefits from the negligible modal cross-talk of a polarisation-maintaining multicore fibre throughout the majority of its length. A tapered region at one end -- known as a photonic lantern -- guides light to a short segment of multimode optical fibre. We show how this fibre enables the projection of high-fidelity structured light to the multimode distal fibre facet via wavefront shaping, providing a level of optical control beyond that possible using conventional multicore fibres which suffer from aliasing due to their spatially separated cores. Furthermore, we demonstrate how high-contrast distal focussing can be re-established, anywhere across the output facet, after the multicore fibre is arbitrarily contorted, without access to the distal end. To achieve this we introduce two {\it in-situ} bend compensation strategies, that rely on a distally mounted guide-star or analysis of light reflected from the distal facet -- offering the future potential for rapid single-shot bend calibration. Our work paves a new route towards the controlled delivery of high-dimensional structured light fields through dynamically flexible optical fibres.
\end{abstract}

\maketitle

Sub-cellular resolution imaging of tissue deep inside the body is a grand challenge in biomedical imaging. Holographic micro-endoscopy is an emerging solution in which images are transported through strands of optical fibre by precisely shaping the propagating optical fields. The narrow footprints of optical fibres -- down to $\sim$100\,$\mu$m in diameter -- have opened up a new minimally-invasive window centimetres deep into living systems~\cite{stibuurek2023110,sun2024lensless}. However, despite promising advances in recent years, high-resolution imaging through {\it flexible} optical fibres, as they are arbitrarily twisted and contorted, remains an open problem. In this work we show how a specially designed type of optical fibre offers a new way to tackle this challenge.

There are two main classes of optical fibre used for micro-endoscopy research: multimode fibres~\cite{cao2023controlling} and multicore fibres~\cite{dremel2026lensless}. Multimode fibres consist of a single core supporting up to thousands of orthogonal spatial modes, where each mode represents an independent information channel. The spatial information carried by light propagating through a multimode fibre is scrambled due to modal dispersion and crosstalk, unrecognisably distorting transmitted images. To overcome this issue, the way in which a multimode fibre transforms light can be characterised by measuring its transmission matrix (TM). The TM represents a digital model of the fibre, predicting how any input field will be transformed when it emerges from the other end of the fibre~\cite{popoff2010measuring}. Once measured, the TM bestows fine control over light transmitted through the fibre using a technique known as wavefront shaping~\cite{vellekoop2007focusing}, enabling diffraction-limited point scanning and advanced imaging modalities through multimode fibres~\cite{gomes2025funnelling,pikalek2025hair,stellinga2021time}.

However, the TM of a multimode fibre is highly sensitive to fibre configuration and changes in an unknown way when the fibre is bent, resulting in the loss of imaging capabilities~\cite{gigan2022roadmap}. To continuously image through such a fibre while it flexes, the TM needs to be somehow recalibrated {\it in-situ}, even while one end of the fibre is embedded in a sample. This is difficult to achieve since TM measurement conventionally requires full optical access to both ends of the fibre. To tackle this problem, several single-ended TM calibration techniques are under development. Approaches include predicting the TM from knowledge of fibre configuration~\cite{ploschner2015seeing}, adaptively switching between multiple pre-calibrated TMs during use~\cite{farahi2013dynamic,wen2023single,wen2025cascaded,mekhail2026portable}, placing spatially varying reflectors or metasurfaces on the output facet~\cite{gu2015design,gordon2019characterizing} and using feedback from distal guide-stars coupled with fibre memory effects~\cite{li2021memory,rosen2015focusing}. Another route being explored is the development of new types of fibre that are more resilient to bending~\cite{flaes2018robustness,turtaev2025deformation}. Despite this progress, imaging through multimode fibres as they are arbitrarily contorted has yet to be achieved.

\begin{figure*}[t]
    \centering
    \includegraphics[width=0.95\linewidth]{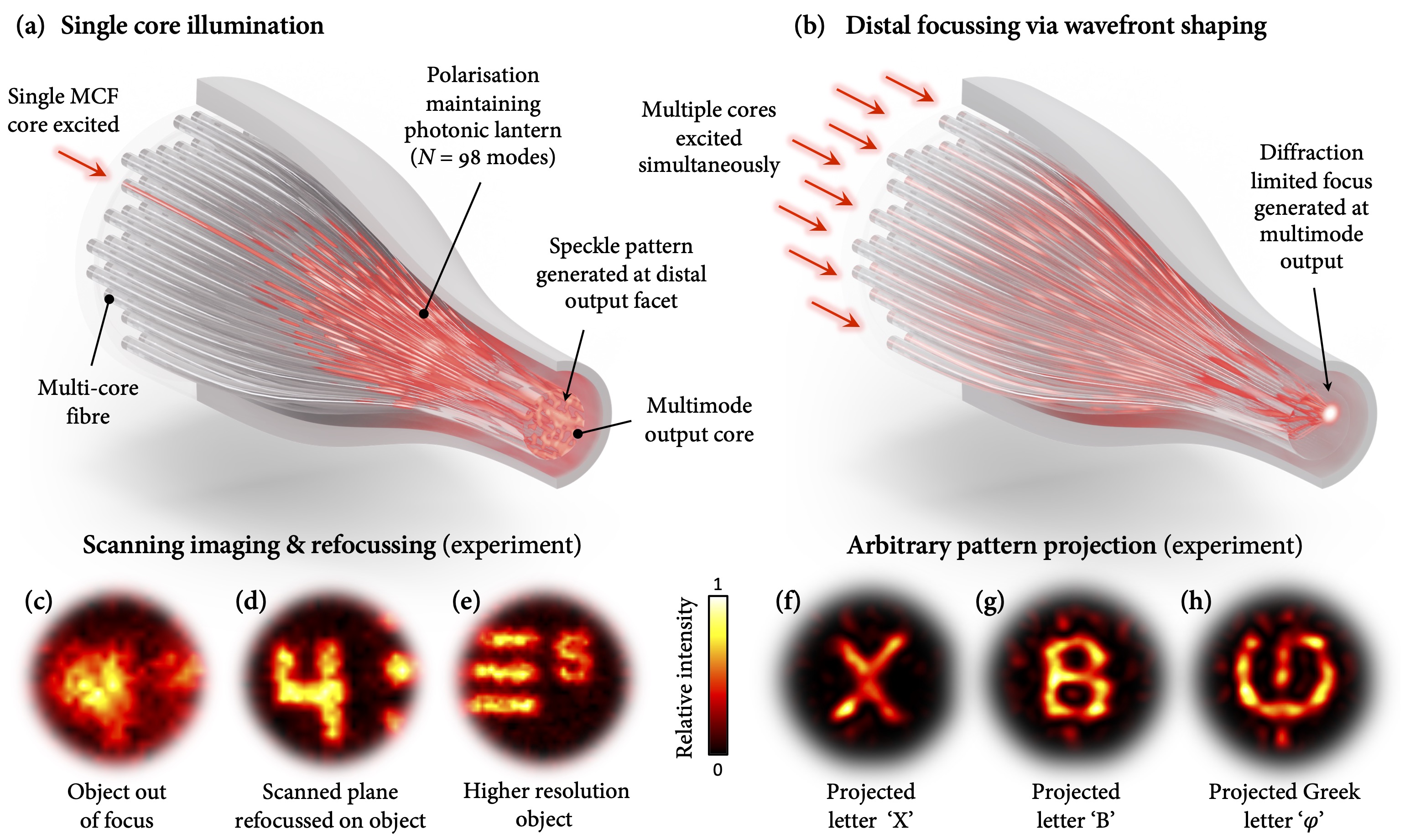}
    \caption{{\bf Imaging and structured light projection through polarisation-maintaining multicore photonic lantern fibres}. (a) Illuminating a single MCF core generates a speckle pattern at the multimode distal facet. (b) Knowledge of the TM of the photonic lantern fibre enables multiple cores to be simultaneously coherently excited with a relative amplitude and phase chosen to generate a high-contrast diffraction-limited focussed spot at a target 3D location at the output. (c-e) Examples of experimentally measured transmission images recorded by scanning a focussed point at the output. The pixel resolution across the field-of-view is limited by the number of MCF cores ($N_{\text{c}} = 98$ in our prototype device). A transmissive calibration slide is placed $\sim80$\,$\mu$m from the image plane of the distal facet. (c) When scanning a focussed spot over the distal facet, the reconstructed image is initially out of focus. (d) The axial plane in which the spots are scanned can be digitally refocussed, thus bringing the object into focus. (e) An image of a different part of the calibration slide showing higher resolution features. (f-h) Beyond focussing, arbitrary pattern projection is also possible.}
    \label{fig:WSPhotonicLantern}
\end{figure*}

The second class of fibre used in micro-endoscopy is multicore fibre (MCF), which guides light through separate single or few mode cores running in parallel. The core-to-core spacing is set to ensure negligible coupling of light between the cores, meaning MCFs have a larger form factor and consequently a lower information density than multimode fibres. Direct imaging through flexible MCFs is achievable if the light transmitted through each core is treated as a single image pixel~\cite{hirschowitz1961endoscopic,polglase2005fluorescence,hsiung2008detection}. This is one approach used in conventional endoscopes, which can also be equipped with distal optics to move the image plane some distance beyond the distal facet, at the cost of further increasing endoscope size. The need for well-separated cores typically constrains the resolution of this direct imaging approach to a few microns at visible wavelengths~\cite{saitoh2016multicore,stone2017low,wood2018high,parker2020core,xu2025high}.

To surpass this resolution limit, wavefront shaping and holographic imaging techniques have also been applied to MCFs~\cite{andresen2013toward,kim2014toward,gordon2019quantitative,badt2022real}. If there is negligible coupling of light between the cores, single-ended calibration of flexible MCFs amounts to the measurement of the additional core-dependent optical phase delays picked up as the fibre bends. For light in a single polarisation state, these core-dependent phase changes can be measured in-situ, without needing access to the distal end~\cite{kuschmierz2018self,choi2022flexible}. Furthermore, the use of helically twisted MCFs has been shown to reduce susceptibility to bending~\cite{tsvirkun2019flexible} and memory-effect based calibration free approaches have also been explored~\cite{stasio2016calibration,porat2016widefield}.

Yet, holographic MCF imaging approaches also face challenges. The separation between the cores means holographic optical control over the output field is under-sampled: there is a mismatch between the number of modes supported by the MCF (equal to the number of cores, $N_{\text{c}}$, when each core supports a single spatial mode) and the larger number of modes needed to fully control light at the distal end of the fibre, $N_{\text{d}}$. This mismatch means that it is not possible to focus all transmitted light at the output using wavefront shaping. The fraction of transmitted optical power that can be shaped as desired, $\eta$, scales according to the mode mismatch ratio: ${\eta =N_{\text{c}}/N_{\text{d}}}$~\cite{vellekoop2007focusing,li2021compressively,gomes2022near}. For typical MCFs, this results in the focus carrying only a few percent of the total transmitted power~\cite{genchi2025two}. The remaining uncontrolled light is scattered to other places in the field-of-view. For regularly spaced cores, this light forms additional unwanted diffraction orders that produce shifted copies of images, reducing the usable field-of-view. Randomly packing the cores blurs out these unwanted diffraction orders into a background signal that reduces the imaging contrast. Thus, while impressive strides forward have been taken in holographic imaging through MCFs, the separation between the cores fundamentally limits the achievable contrast and field-of-view.

In this work we demonstrate how a bespoke optical fibre architecture, blending attributes of both MCFs and multimode fibres, offers a potential way forward to overcome these challenges. We employ a polarisation-maintaining MCF that has been tapered into a short segment of multimode fibre at one end via a photonic lantern~\cite{birks2015photonic,fontaine2022photonic}. Figure~\ref{fig:WSPhotonicLantern} shows a schematic of this photonic lantern transition. The taper deterministically couples light from the individual single-mode MCF cores into the multimode waveguide. 

Crucially, the photonic lantern is engineered such that the number of modes supported by the multimode waveguide is matched to the number of single mode MCF cores. This ensures wavefront shaping bestows close to full control over the field projected to the distal output, enabling high contrast imaging and eliminating the aliasing issues that occur with conventional MCFs. In this article we henceforth refer to this fibre as a {\it photonic lantern fibre}. In proof-of-principle experiments, we demonstrate high contrast diffraction-limited scanning imaging through this photonic lantern fibre. Furthermore, we show how this fibre admits in-situ bend compensation: when the MCF bends, the full TM of the fibre can be re-calibrated using measurements made with access only to the proximal fibre facet.\\

\noindent{\bf Photonic lantern fibre fabrication and characterisation}

\noindent We fabricate our photonic lantern fibre from a $\sim$1\,m length of bespoke polarisation-maintaining MCF with $N_{\text{c}}=98$ cores stacked in a rectangular lattice~\cite{stone2014highly}. The polarisation-maintaining nature of this MCF is crucial for our proposed in-situ bend compensation approaches, as explained in more detail below. At a wavelength of $\lambda = 633$\,nm, each MCF core supports a pair of orthogonal plane-polarised versions of a single spatial mode. If the linear input polarisation is aligned along one of the two orthogonal principal axes (which are oriented along columns/rows of the cores in the MCF~\cite{stone2014highly}), the input polarisation state is maintained throughout propagation along the cores.

\begin{figure*}[t]
    \centering
    \includegraphics[width=1\linewidth]{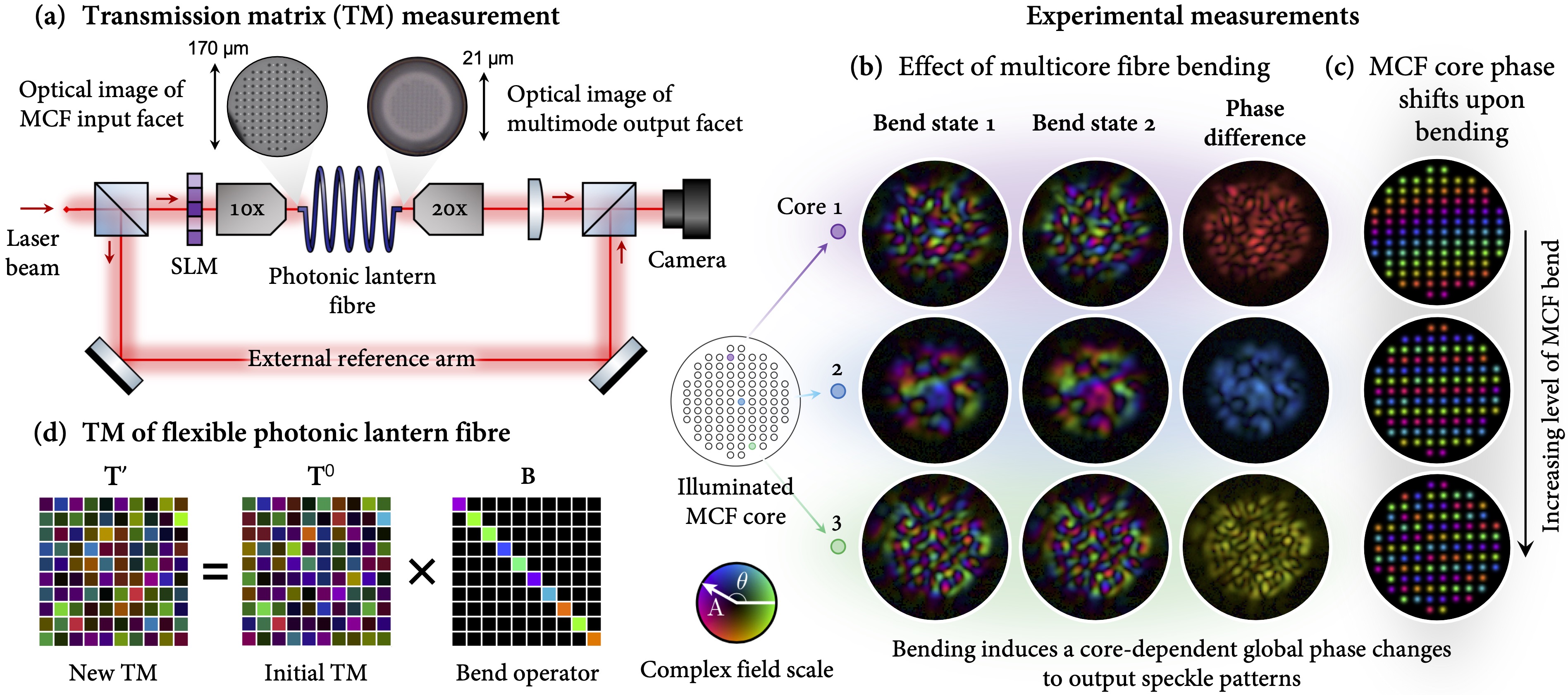}
    \caption{{\bf The effect of multicore fibre bending on the TM of the photonic lantern fibre}. (a) A simplified schematic showing the photonic lantern fibre TM measurement set-up. See Methods \S1 for details. Upper insets show optical microscope images of the input MCF (left) and output multimode (right) facets of the photonic lantern fibre. Within the image of the MCF facet, white disks correspond to cores and dark disks correspond to stress rods that induce birefringence for polarisation-maintaining operation. (b) Experimental measurements of the output speckle patterns measured when individually illuminating $3$ different cores (rows) with the MCF held in two different bend states (first $2$ columns). The `Illuminated MCF core' key indicates which core was excited in each case. The phase difference between the speckle patterns transmitted in each MCF bend state are plotted as a function of location across the distal facet. Here the amplitude of these plots is given by the mean amplitude of the speckle patterns in bend states $1$ and $2$. We observe well correlated output complex speckle patterns across the two bend states, with a uniform core-dependent phase difference across the output facet. (c) Experimentally measured relative MCF core phase shifts for an increasing degree of MCF bending. Here the phase shift of the $n^{\text{th}}$ core is given by ${\delta\phi_n = \text{arg}[(\Vecs_n^{(1)})^\dagger\cdot\Vecs_n^{(2)}]}$, where $\Vecs_n^{(i)}$ is the vectorised speckle pattern emitted from the $n^{\text{th}}$ core and $i$ denotes bend state $1$ or $2$. (d) The TM of a flexible photonic lantern fibre is well-approximated as the product of two matrices: a fixed matrix $\VecT^0$ associated with the transformation enacted by the photonic lantern transition, and a unitary diagonal `bend operator' $\VecB$ which is an MCF configuration-dependent unitary matrix (Eqn.~\ref{eq:bendOp}). Here we depict an $N=10$ mode illustration for clarity.}
    \label{fig:TMBending}
\end{figure*}

A photonic lantern is created at one end by inserting the MCF into a Fluorine-doped (F-doped) jacket and tapering this assembly over a distance of $\sim4$\,cm to a multimode waveguide. By adjusting the stretching rate and temperature during tapering, the multimode waist diameter can be tailored to the right size to support approximately the same number of spatial modes as the number of modes supported by the MCF -- here we realise a core diameter of $d = 21$\,$\mu$m. During this tapering process, the F-doped jacket collapses onto the MCF and forms the lower refractive index cladding, while the entire MCF forms the core of the output multimode waveguide. The multimode segment is then cleaved, yielding two polarisation-maintaining photonic lantern fibres. Figure~\ref{fig:TMBending}(a) shows optical images of the input MCF facet and output multimode facet of the fibre.

We experimentally measure the numerical aperture (NA) of the output multimode waveguide to be ${\text{NA}\sim0.2}$ (see Methods \S2). For a step-index multimode fibre, the number of spatial modes per polarisation state, $N_\text{d}$, is given by ${N_\text{d}\sim [\pi d\,\text{NA}/(2\lambda)]^2}$, resulting in ${N_\text{d}\sim110}$ in our case. Thus, using our photonic lantern fibre, we estimate that the maximum fraction of transmitted optical power that can -- in principle -- be shaped as desired is ${\eta = N_\text{c}/N_\text{d}\sim0.9}$, indicating that high-contrast focussing and imaging should be achievable.

We next characterise the level of polarisation coupling within the photonic lantern fibre: each core is illuminated in turn with linearly polarised light aligned with one of the principal axes of the MCF, and the fraction of transmitted optical power coupled into the orthogonal polarisation state is measured. We find the level of polarisation coupling to be ${c_\text{pol}\sim6\%}$, averaged over all cores. As there is negligible polarisation coupling within the MCF itself~\cite{stone2014highly}, the measured value of $c_\text{pol}$ is almost entirely due to propagation through the photonic lantern transition. Nonetheless, this relatively low level of polarisation coupling suggests that a high degree of optical control can be achieved with scalar wavefront shaping using uniformly linearly polarised input fields.\\

\noindent{\bf Wavefront shaping through photonic lantern fibres}

\noindent To experimentally investigate focussing and light control through our photonic lantern fibre, we build an optical setup capable of measuring its TM and implementing wavefront shaping~\cite{lp2020coherent}. Figure~\ref{fig:TMBending}(a) shows a simplified schematic (see also Methods \S1 and Supplementary Information \S1). A spatial light modulator (SLM) is used to shape input laser light, allowing excitation of individual or combinations of MCF cores with beams of independently tunable intensity and phase. In these experiments, diffractive beam shaping is achieved with a high-speed Digital Micro-mirror Device (DMD)~\cite{lee1979binary,turtaev2017comparison}. The resulting complex optical fields generated at the multimode output of the photonic lantern are measured using phase shifting holography (see Methods \S1).

Light propagating along a single core of the MCF spreads across many multimode waveguide modes upon passage through the tapered photonic lantern transition (see Fig.~\ref{fig:WSPhotonicLantern}(a)), thus generating a speckle pattern at the multimode output facet -- experimental examples are shown in Fig.~\ref{fig:TMBending}(b). In this way, the photonic lantern performs an (approximately) unitary spatial basis transformation, linearly mapping the $N_{\text{c}}$ orthogonal single modes of the MCF to a set of $N_{\text{c}}$ orthogonal speckle patterns supported by the multimode waveguide (sometimes referred to as the `principal lantern modes'~\cite{taras2026empirical}). 

The TM represents a digital model describing how any monochromatic input field, $\Vecu$, will evolve into an output field $\Vecv$, where $\Vecv = \VecT\cdot\Vecu$. Here $\Vecu$ is an $N_{\text{c}}$-dimensional column vector representing the input field in the MCF core basis, in which the $n^{\text{th}}$ complex element signifies the amplitude and phase of light illuminating core $n$. $\Vecv$ is an $M$-dimensional column vector representing the pixelated field at the multimode output end in a vectorised format, where $M$ is equal to the number of camera pixels used to record the field. Typically $M>N_{\text{c}}$ in order to ensure the output field is well sampled.

Once measured, the TM reveals the required input field, ${\Vecu = \Vecu_{\text{t}}}$, that will propagate through the photonic lantern fibre to generate a desired target distal field ${\Vecv = \Vecv_{\text{t}}}$ (within the spatial bandwidth of the fibre) at the multimode output:
\begin{equation}\label{eqn:TM_invert}
    \Vecu_{\text{t}} = \VecT^{+}\cdot\Vecv_{\text{t}},
\end{equation}
where $\VecT^{+}$
is the Moore-Penrose pseudo-inverse of the (potentially non-square) matrix $\VecT$. As calculation of the pseudo-inverse is highly sensitive to measurement noise in $\VecT$, in this work we use the approximation ${\VecT^{+}=\VecT^{\dagger}}$, where $^\dagger$ represents the conjugate transpose operation, which is a reasonable assumption when the TM is expected to be close to unitary, as in this case. To focus onto the position of the $m^{\text{th}}$ camera pixel at the output (as depicted in Fig.~\ref{fig:WSPhotonicLantern}(b)), we set $\Vecv_{\text{t}}$ to a one-hot encoded vector in which element $m$ is set to 1, while all other elements are set to 0. Inserting this into Eqn~\ref{eqn:TM_invert}, along with knowledge of the pre-measured TM, yields the complex elements of $\Vecu_{\text{t}}$, whose elements correspond to the amplitude and phase with which we should simultaneously excite each of the MCF cores at the proximal end. 

Figure~\ref{fig:WSPhotonicLantern}(c-h) shows experimental examples of transmission imaging and pattern projection through the photonic lantern fibre. Once the TM of the photonic lantern is known, the image plane can be digitally refocused to any distance from the fibre tip as shown in Fig.~\ref{fig:WSPhotonicLantern}(c-d). This is achieved by adjustment of the input excitation field via incorporation of a free-space propagation operator into Eqn.~\ref{eqn:TM_invert} -- see Methods \S3.\\

\noindent{\bf The effects of multicore fibre bending}

\noindent Accurate knowledge of the fibre's TM is crucial for control of optical fields at the output. Yet the TM is highly sensitive to the bend state of the MCF. Figure~\ref{fig:TMBending}(b) demonstrates how propagating fields are modified when the MCF is bent. Since light does not couple between the cores, or between the two polarisation states within each individual core, we see that when only a single core is illuminated, the spatial structure of each output speckle pattern remains unchanged upon bending of the MCF. However, the optical path length of each core is altered in a core-dependent manner, and so the output speckle pattern generated by each core experiences a different global phase shift $\delta\phi_n$ (where $n$ indexes the core). Figure~\ref{fig:TMBending}(c) shows the relative changes in the global phase of light propagating along all $98$ cores as the severity of MCF bend is progressively increased.

\begin{figure*}[t]
    \centering
    \includegraphics[width=1\linewidth]{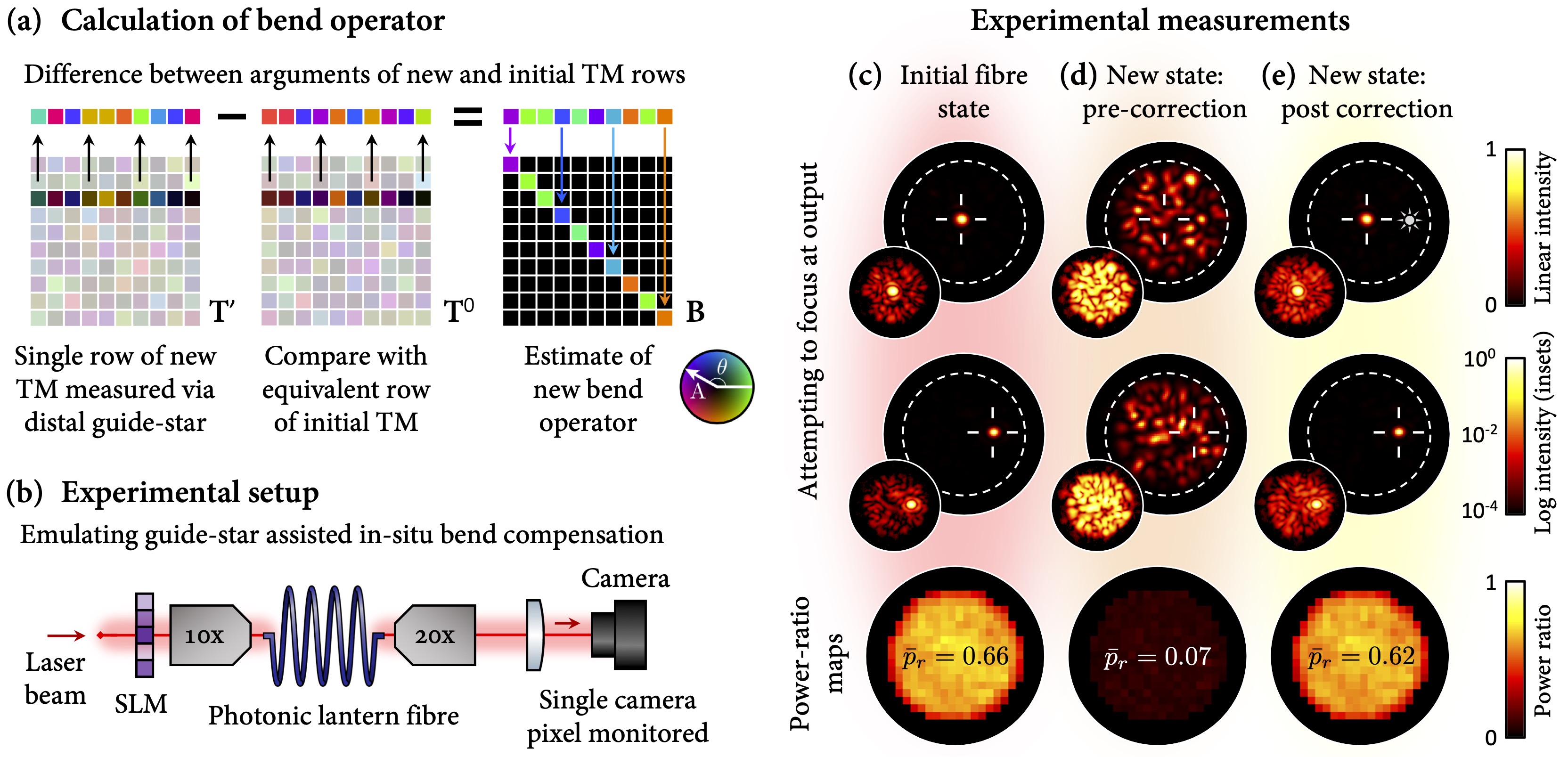}
    \caption{{\bf Emulation of guide-star assisted in-situ bend compensation}. (a) A graphical depiction of bend operator calculation. The input field that focusses onto the guide-star, which can be found with only proximal end access, reveals a single row of the new TM $\VecT'$. The difference in the phase of the elements of this row and the equivalent row in the original TM $\VecT^0$, reveals the bend operator $\VecB$. (b) A simplified schematic showing the experimental set-up to emulate guide-star assisted bend compensation. The signal from a single camera pixel imaging the output facet of the photonic lantern fibre is used to emulate the proximal signal from a guide-star mounted on the same location on the fibre tip. (c-e) Control over the propagation of light through photonic lantern fibres under different MCF bend conditions. The upper two rows show examples of distal focused spot generation when the MCF is: (c) in its initial state; (d) after the MCF is bent into a new state but prior to bend compensation; and (e) with the MCF in its new bend state with guide-star assisted bend compensation applied. The target spot location is highlighted with a cross-hair. The location of the guide-star pixel on the distal facet is indicated by a grey star in the top panel of (e). The Insets to each panel show the projected intensity on a log scale. The bottom row shows a power ratio map for each case, indicating how well it is possible to focus to anywhere across the distal fibre facet. We see that after the MCF is bent to a new state, applying guide-star assisted bend correction recovers the ability to focus anywhere on the distal facet with high fidelity.
    }
    \label{fig:guide-star}
\end{figure*}

Assuming there is negligible polarisation coupling within the MCF or the photonic lantern, the optical system can be accurately represented in the scalar approximation. If the photonic lantern transition itself remains unperturbed, bending the MCF results in unknown different global phase shifts imparted to each column of the scalar TM describing the transport of a single polarisation state through the optical system. In this case, the TM of the fibre in its new state, $\VecT'$, is related to the original TM, $\VecT^0$, by
\begin{equation}\label{eq:bendOp}
    \VecT' = \VecT^0\cdot\VecB,
\end{equation}
where we refer to matrix $\VecB$ as the `bend operator': an $N_{\text{c}}\times N_{\text{c}}$ diagonal unitary matrix holding along its diagonal the global phase changes, experienced by light in each core, between the initial and new MCF configuration (see Fig.~\ref{fig:TMBending}(d)). The $n^{\text{th}}$ element along the diagonal of $\VecB$ is given by ${B_{n,n} = \exp(i\delta\phi_n)}$, where $\delta\phi_n$ is the phase shift imparted to light in core $n$ when the MCF bends. Without knowledge of $\VecB$, control of light at the multimode output is lost, and thus so are imaging capabilities. To rectify this situation, our task is to measure and compensate for these core-dependent phase shifts without relying on access to the distal facet to remeasure the TM. In what follows we demonstrate two different approaches to finding this bend operator.\\

\noindent{\bf Guide-star assisted in-situ bend compensation}

\noindent Our first proposed strategy involves placing a `guide-star' on the distal photonic lantern fibre tip. The concept of aberration correction using a guide-star was originally developed in astronomy (from where the name stems), and more recently extended to microscopy. In microscopy applications, a guide-star is a point within an aberrating medium capable of signalling the local intensity of the illuminating field to the outside world -- through, for example, fluorescence emission~\cite{horstmeyer2015guidestar}. Guide-stars permit focussing and imaging through scattering media without direct optical access to the image plane. We recently showed how to generalise the application of guide-stars to partially image through flexible multimode fibres~\cite{li2021memory}. This approach enables estimation of the TM of a multimode fibre without optical access to its distal end, using the light radiating from a fluorescent guide-star mounted on the fibre tip, combined with an estimate of the eigenbasis of the multimode fibre's TM~\cite{ploschner2015seeing}.

Here we describe how a guide-star can similarly be used to recover the bend operator $\VecB$ of a polarisation-maintaining multicore photonic lantern fibre, while only requiring optical access to its proximal end. 
To proceed, we first measure the TM of the fibre in its initial bend state, $\VecT^0$, using the setup shown in Fig.~\ref{fig:TMBending}(a). This pre-calibration step is carried out with full optical access to the output end of the photonic lantern. The fibre can then be deployed into a sample, which will typically change its bend state and mean that optical access to the output end is no longer available.

With the MCF in a new bend state, we next determine the input field $\Vecu_\text{gs}$ required to focus onto a distal guide-star~\cite{horstmeyer2015guidestar}. This can be achieved in-situ, with access only to the proximal end of the photonic lantern fibre, by analysing the fluorescent particle's emission when illuminated with different input states of light. For example, a focus on the guide-star gradually forms by iteratively optimising the phase of the fields illuminating each MCF core to maximise the total fluorescent signal received back at the proximal end of the fibre~\cite{vellekoop2007focusing,conkey2012genetic}. Alternatively, $\Vecu_\text{gs}$ can be determined by phase shifting holography~\cite{vcivzmar2010situ,conkey2012high}, which is the approach we use here (see Methods \S4). 

If the guide-star is located at pixel $m_0$ on the output plane, the intensity arriving at this point is given by ${I_{m_0} = \left|\Vect_{m_0}^\top\cdot\Vecu_\text{gs}\right|^2}$, where $\Vect_{m_0}^\top$ specifies row $m_0$ of $\VecT'$ (here $^\top$ represents the transpose operation). From this we can see that $I_{m_0}$ is maximised when $\Vect_{m_0} = \Vecu_\text{gs}^*$. Thus, by focussing on the guide-star, we have found the $m_0^{\text{th}}$ row of the $\VecT'$. To recover all rows of $\VecT'$ from this information, we can see from Eqn.~\ref{eq:bendOp} that since $\VecB$ is diagonal, it follows that
\begin{equation}\label{eq:transEl}
    T'_{m_0,n} = T^0_{m_0,n}\times B_{n,n}.
\end{equation}
Here subscripts index matrix elements by row and column. Rearranging Eqn.~\ref{eq:transEl} for $B_{n,n}$ and taking the argument of both sides, noting that ${\arg(B_{n,n}) = \delta\phi_n}$, yields an equation for the phase of the $n^{\text{th}}$ diagonal element of the bend operator~\cite{li2021memory}:
\begin{equation}\label{eq:reflPhi}
    \delta\phi_n = \arg\left(T'_{m_0,n}\right) - \arg\left(T^0_{m_0,n}\right).
\end{equation}
This process is shown schematically in Fig.~\ref{fig:guide-star}(a). Once the bend operator $\VecB$ has been measured, the TM of the photonic lantern fibre in its new bend configuration, $\VecT'$, can then be calculated using Eqn.~\ref{eq:bendOp}.

We now experimentally test our guide-star assisted bend compensation strategy. We conduct proof-of-principle experiments by imaging the output facet of the photonic lantern using a setup shown in Fig.~\ref{fig:guide-star}(b) and analysing the signal from a single camera pixel. In this way we emulate the signal obtainable by collecting (at the proximal end of the fibre) the emission from a fluorescent guide-star mounted on the distal fibre tip.

To provide a baseline, in Fig.~\ref{fig:guide-star}(c) we first demonstrate the fidelity of output light control with full-knowledge of the TM $\VecT^0$ before the fibre is bent. Here the upper and middle panels show examples of focussing to two different output locations. To quantify the focus fidelity we measure the experimental {\it power-ratio}, $p_r$ $(0 \le p_r \le 1)$, given by the optical power focussed into a target spot on the distal facet divided by the total transmitted optical power passing through the entire output fibre core area. We raster scan a focus over the output facet and calculate a power-ratio `map': the power-ratio as a function of focus position, which is shown in the bottom panel in Fig.~\ref{fig:guide-star}(c). From this map we calculate the mean power-ratio, averaged over position, given by $\bar{p}_r = 0.66$. This is substantially higher than achievable if focussing through the unmodified MCF, and may be further improved towards our maximum estimate of ${\eta\sim0.9}$ by improving system stability, reducing TM measurement noise, and optimising hologram design~\cite{gomes2022near}.

The MCF is now bent into a new configuration. If we continue to apply the originally measured TM $\VecT^0$, this change of fibre state completely disrupts focussing capabilities, as shown in Fig.~\ref{fig:guide-star}(d): we observe the output foci are obliterated into speckle patterns, and the mean power ratio falls to $\bar{p}_r = 0.07$. To recover control of the light field at the distal facet, we now follow the protocol described above to measure the bend operator.

To find the input field that focusses on the guide-star, we illuminate each core in turn, and carry out phase shifting holography with a reference beam co-propagating through one of the cores (see Methods \S4). From these measurements we recover row $m_0$ of the new TM $\VecT'$. We apply Eqn.~\ref{eq:reflPhi} to recover the bend operator $\VecB$ and use it to calculate the entire new TM $\VecT'$ via Eqn.~\ref{eq:bendOp}. Using $\VecT'$ to control light through the fibre in its new bend state, we see that focussing capabilities across the entire distal facet are restored, as shown in Fig.~\ref{fig:guide-star}(e). The mean power ratio using the corrected TM is $\bar{p}_r = 0.62$.

This is an encouraging result, however guide-star assisted bend compensation relies on the output facet of a photonic lantern fibre being engineered to incorporate a fluorescent particle -- one that must be suitably encapsulated so that it is not accidentally removed during use. Therefore, we next explore an approach that circumvents this requirement.\\

\noindent{\bf Reflection-based in-situ bend compensation}

\noindent Our second bend compensation strategy relies on decoding the core-dependent phase shifts $\delta\phi_n$ from light reflected from the refractive index interface at the distal multimode facet. We first consider the spatial transformation undergone by reflected light when a single MCF core is illuminated. As discussed above, single core excitation generates a unique speckle pattern at the multimode waveguide output (Fig.~\ref{fig:TMBending}(b)). Upon reflection from the distal interface, this speckle pattern excites many MCF cores on the return path through the photonic lantern (note: for the avoidance of confusion, for reflected light to instead follow back along {\it the same} path as the outbound excitation, phase conjugation at the reflective interface would be required).

\begin{figure*}[t]
    \centering
    \includegraphics[width=1\linewidth]{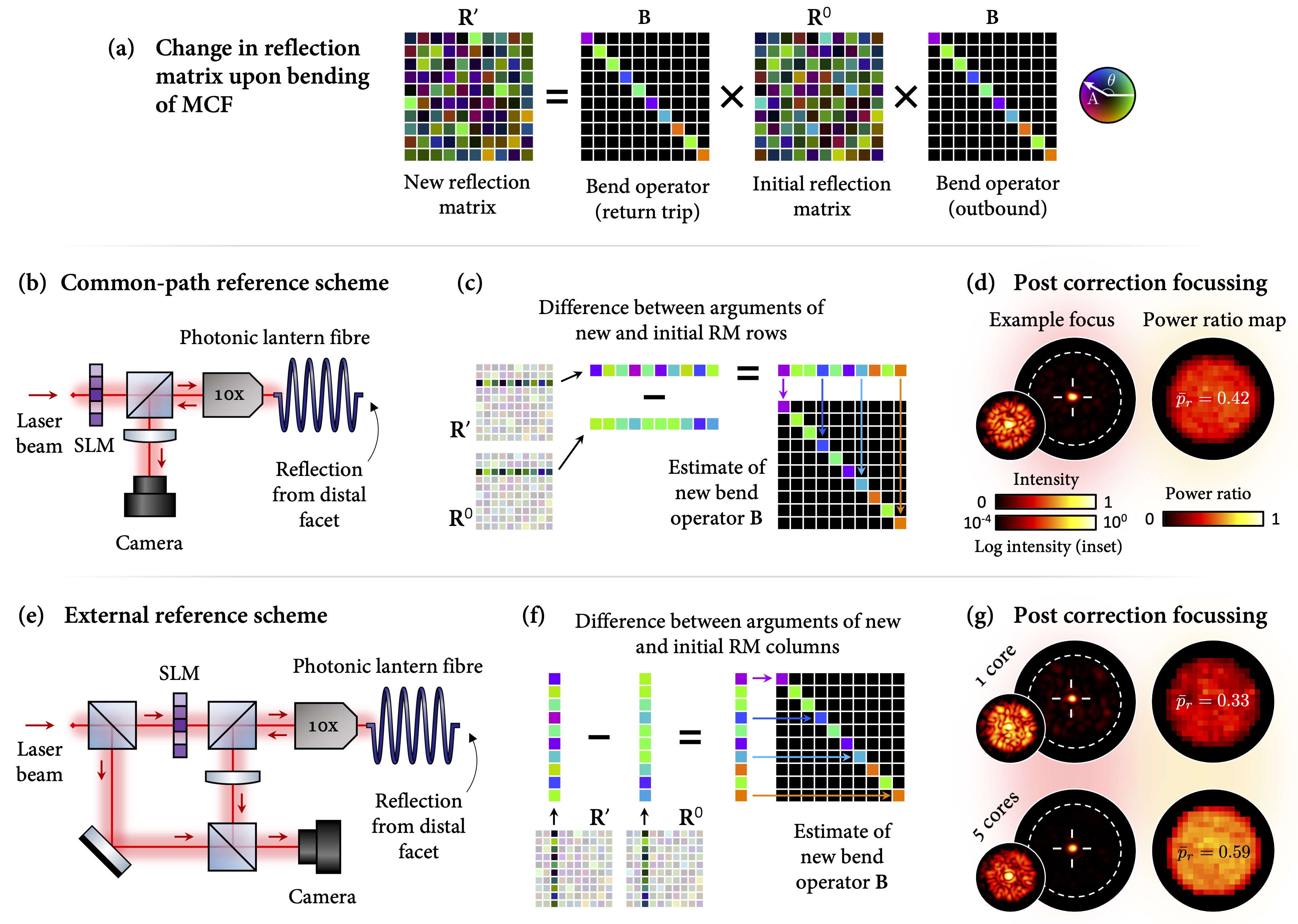}
    \caption{{\bf Reflection matrix based in-situ bend compensation}. (a) Relationship between the reflection matrix of a photonic lantern fibre in two different bend states (Eqn.~\ref{eq:reflNew}). (b) Simplified schematic showing experimental setup for common-path referenced reflection based bend compensation (also see Methods \S5 and Supplementary Information \S1). (c) Recovery of bend operator from measurement of a single row of $\VecR^0$ and $\VecR'$ (Eqn.~\ref{eq:reflPhiRow}). In practice, we are able to simultaneously retrieve the phase differences of nearly all new and initial reflection matrix rows, and average these measurements to enhance the SNR -- see Methods \S5. (d) Experimentally measured focussing fidelity after common-path reference reflection based bend compensation is applied. The left panel shows an example of a distal focus. A log scale plot of the same data is shown inset. The right panel is a power-ratio map indicating the focal quality across the output facet. (e) Simplified schematic showing experimental setup for externally referenced reflection based bend compensation (see Methods \S6 and Supplementary Information \S1). (f) Recovery of bend operator from measurement of a single column of $\VecR^0$ and $\VecR'$ (Eqn.~\ref{eq:reflPhiCol}). (g) Experimentally measured focussing fidelity after external reference reflection based bend compensation is applied. The left panels show examples of distal foci with log scale plots of the same data inset. Right panel is power-ratio map. The upper row shows the focussing achieved after probing a single MCF core, the lower row shows that achieved after probing $5$ cores and combining these measurements. 
    }
    \label{fig:reflection}
\end{figure*}
By sequentially illuminating each MCF core in turn, and holographically measuring the optical field of light reflected back out of each MCF core, the reflection matrix of the photonic lantern fibre in its original configuration, $\VecR^0$, can be measured:
\begin{equation}\label{eq:reflOrig}
\VecR^0 = \VecT^{0\top}\cdot\VecH\cdot\VecT^0.
\end{equation}
Here $\VecT^0$ is the outbound TM (as above), matrix $\VecH$ accounts for the Fresnel reflection at the distal interface and any deviation of the output facet from a flat plane perpendicular to the optical axis, and 
$\VecT^{0\top}$ is the TM of the fibre when propagating from the distal facet back to the proximal end.

When the MCF bends, the reflection matrix of the photonic lantern fibre in the new state, $\VecR'$, is given by
\begin{equation}\label{eq:reflNew}
\begin{split}
    \VecR'& = \VecT'^{\top}\cdot\VecH\cdot\VecT'\\
    & = \left(\VecT^0\cdot\VecB\right)^\top\cdot\VecH\cdot\left(\VecT^0\cdot\VecB\right)\\
    & = \VecB\cdot\VecR^0\cdot\VecB,
    \end{split}
\end{equation}
where $\VecT'$ is the outbound TM of the photonic lantern fibre in its new state, as defined in Eqn.~\ref{eq:bendOp}. Light propagating along each core picks up a phase shift on both the outbound and return trips through the optical system, and so we see that applying the bend operator on either side of the original reflection matrix yields the new reflection matrix (see Fig.~\ref{fig:reflection}(a)).

Since $\VecB$ is diagonal, Eqn.~\ref{eq:reflNew} shows that the element of $\VecR'$ in row $n$ and column $m$ is given by 
\begin{equation}\label{eq:refEl}
    R'_{n,m} = B_{n,n}\times R^0_{n,m}\times B_{m,m},
\end{equation}
where $m$ indexes the input MCF core, and $n$ indexes the output MCF core. Rearranging Eqn.~\ref{eq:refEl} for $B_{m,m}$ and taking the argument of both sides yields:
\begin{equation}\label{eq:reflPhiRow}
    \delta\phi_m = \arg\left(R'_{n,m}\right) - \arg\left(R^0_{n,m}\right) - \delta\phi_n,
\end{equation}
where ${\arg(B_{m,m}) = \delta\phi_m}$. Equation~\ref{eq:reflPhiRow} shows that the measurement of a single {\it row} of $\VecR^0$ and $\VecR'$ (row $n$) is sufficient to recover all elements of the bend operator $\VecB$, up to a global phase factor of $\delta\phi_n$. This is shown schematically in Fig.~\ref{fig:reflection}(c).

Alternatively, rearranging Eqn.~\ref{eq:refEl} for $B_{n,n}$ and taking the argument of both sides yields a subtly different equation:
\begin{equation}\label{eq:reflPhiCol}
    \delta\phi_n = \arg\left(R'_{n,m}\right) - \arg\left(R^0_{n,m}\right) - \delta\phi_m,
\end{equation}
which shows that measuring a single {\it column} of $\VecR^0$ and $\VecR'$ (column $m$) is also sufficient to recover all elements of the bend operator $\VecB$, up to a global phase factor of $\delta\phi_m$ (see Fig.~\ref{fig:reflection}(f)). Therefore, measurement of either a row or column of the reflection matrices yields the same information -- which is apparent since both $\VecR^0$ and $\VecR'$ are symmetric matrices. These alternatives correspond to two different experimental approaches to achieve in-situ bend compensation: holographic reflection measurements using (i) a `common-path' reference beam (row measurement) or (ii) an `external' reference beam (column measurement). We now explore the implementation of these strategies experimentally.\\

\noindent(i) {\it Common-path referenced reflection bend compensation}

\noindent The first reflection-based bend compensation strategy relies on probing one row of the reflection matrix of the photonic lantern fibre~\cite{milne2023coherent}. As before, we first measure the forward TM of the photonic lantern fibre in its initial state, $\VecT^0$. We also measure one row of the reflection matrix with the MCF in its initial state, using an experimental setup shown in Fig.~\ref{fig:reflection}(b) (also see Supplementary Information \S1). The photonic lantern fibre is probed in the MCF core basis by exciting each core sequentially in a pair-wise fashion with a common reference core, which provides a common-path reference beam (i.e., a reference beam that also co-propagates through the photonic lantern fibre -- see Methods \S5). Probing core $m$, and observing the light emanating back out of core $n$, while phase stepping the reference beam, measures $R^0_{n,m}$, and we probe all cores to obtain row $n$ of $\VecR^0$.

We now assume the fibre is deployed in a sample and access to the output is no longer available. We make a second set of reflection measurements with the MCF in a new state, yielding row $n$ of $\VecR'$. Applying Eqn.~\ref{eq:reflPhiRow} recovers the bend operator (see Fig.~\ref{fig:reflection}(c)). In principle, it is only necessary to observe light reflected back along a single detection MCF core. However for each probe measurement it is possible to simultaneously monitor light reflected back through all cores using the proximal camera. Therefore, we combine such measurements to enhance the signal to noise ratio (SNR) -- see Methods \S5.

Figure~\ref{fig:reflection}(d) shows an experimental demonstration of in-situ bend compensation using this common-path reference scheme. We find the mean power ratio when focussing through the photonic lantern fibre using the corrected TM is $\bar{p}_r = 0.42$. \\

\noindent(ii) {\it Externally referenced reflection bend compensation}

\noindent Our first two bend compensation methods require $N_{\text{c}}$ sequential measurements to recover the bend operator. This may limit correction rates if scaling up these approaches to thousands of cores. Our final in-situ bend compensation method -- corresponding to measurement of a single column of the reflection matrix -- circumvents this issue: in principle all elements of the bend operator can be measured simultaneously with a single probe measurement.

Figure~\ref{fig:reflection}(e) shows the experimental setup in this case (also see Supplementary Information \S1). As before, we first characterise the forward TM of the photonic lantern fibre in its initial state, $\VecT^0$. We also measure the $m^{\text{th}}$ column of the reflection matrix $\VecR^0$: core $m$ is illuminated, and the reflected optical field emanating from all MCF cores is measured simultaneously, by observing the interference of this light with a coherent external reference beam. Here we use phase shifting holography with $4$ separate measurements, although single-shot off-axis digital holography is possible in principle~\cite{cuche2000spatial}. Repeating this reflection measurement with the MCF in a new state yields the $m^{\text{th}}$ column of $\VecR'$. The bend operator is then found by applying Eqn.~\ref{eq:reflPhiCol} (see Fig.~\ref{fig:reflection}(f)).

The upper $2$ panels of Fig.~\ref{fig:reflection}(g) show a demonstration of in-situ bend compensation using this strategy. We first reconstruct $\VecB$ from the measurement of a single column of $\VecR^0$ and $\VecR'$ as described above. However, in this case, we find that the light emanating from some of the MCF cores is obscured by residual stray excitation light. For an input intensity of $I_{\text{in}}$ focussed into a single core, the average intensity of the signal emanating back out of each core is roughly ${I_{\text{out}}= I_{\text{in}}\times r/ N_{\text{c}} \sim I_{\text{in}}\times3\text{e}^{-4}}$, where ${r\sim 0.03}$ is the reflection coefficient of the glass-air interface at the distal fibre facet. Excitation light is shaped by a DMD situated in the Fourier plane of the input MCF facet. When exciting a single core, a small percentage of light from the DMD is also diffracted into higher orders which reflect from the MCF input facet to the detection camera. Even though these higher order diffraction orders from our illumination system are very weak, they are enough to overwhelm the relatively small signal reflected back along some of the MCF cores. This limits the number of diagonal elements of $\VecB$ we are able to accurately recover, and consequently limits the mean power-ratio using the bend-corrected TM to $\bar{p}_r = 0.33$.

To overcome this issue we sequentially illuminated $5$ different cores (thus varying the subset of obscured cores in each case) and combined the results to obtain a reflected signal from all cores. This approach improved the mean power-ratio to $\bar{p}_r = 0.59$, as shown in the lower $2$ panels of Fig.~\ref{fig:reflection}(g). In the future this technical issue may be mitigated by cleaving or polishing the input end of the MCF at a small angle, thus redirecting reflected light away from the detection camera, or using a small index matched wedge prism as described in ref.~\cite{lee2020reciprocity}. Using a more efficient high-speed beam shaper~\cite{rocha2024fast} and positioning it close to the image plane of the input facet~\cite{milne2023coherent} will also help to reduce the effects of unwanted scattered light.\\

\noindent{\bf Bend sensitivity of the photonic lantern transition itself}

\noindent Our bend compensation strategies rely on the assumption that only the MCF undergoes changes in configuration, while the photonic lantern transition remains rigidly fixed. We now explore the importance of this criterion by studying the sensitivity of our system to bending of the photonic lantern itself. In our experiments, the photonic lantern is held rigid within a glass capillary, so here we investigate its bend sensitivity in simulations, using the finite difference-beam propagation method~\cite{pedrola2015beam} -- see Methods \S7 for details of our custom-written model.

We simulate light propagation through a photonic lantern with equivalent parameters to our experiment, and observe how the TM of this photonic lantern changes as it is gradually bent with a uniform radius of curvature. We apply our guide-star and reflection-based bend compensation strategies to attempt to control the evolution of light through the bent photonic lantern. Evidently these strategies operate under the assumption that only global phase changes to the transmitted speckle patterns of the speckle basis occur -- so any such changes induced by bending the photonic lantern will be corrected by our methods. Yet bending of the photonic lantern also induces changes in the shape of the transmitted speckle patterns, which our methods will fail to account for. We use our model to explore to what extent transmitted speckle patterns decorrelate as the photonic lantern is bent, resulting in loss of control over the output field.

\begin{figure}[t]
    \centering
    \includegraphics[width=1\linewidth]{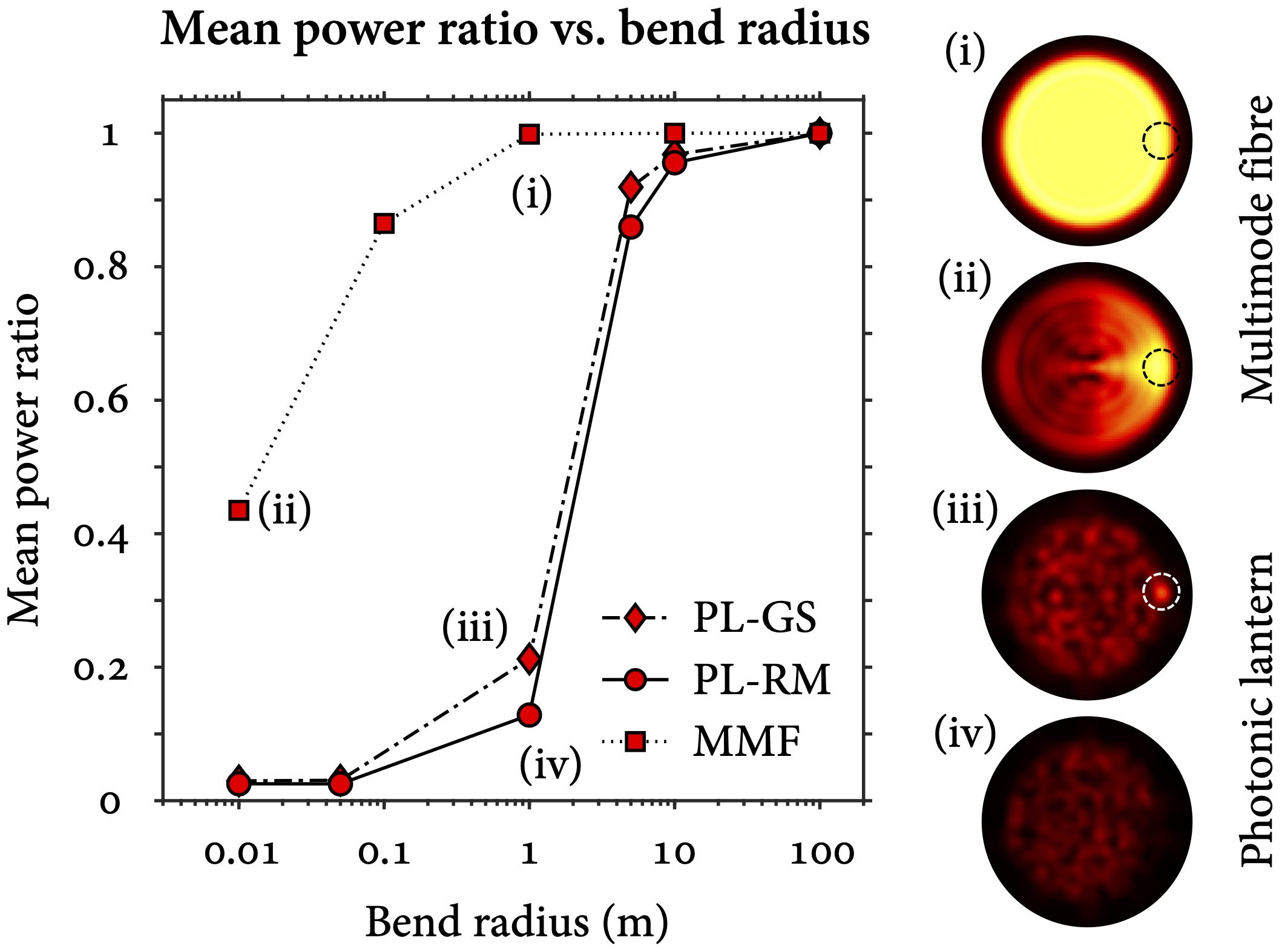}
    \caption{{\bf Simulated bend sensitivity of photonic lantern transition}. The graph shows a comparison of the mean power ratio as a function of bend radius for a photonic lantern bent while guide-star based correction is applied (PL-GS, diamonds), and a photonic lantern bent while reflection matrix based correction is applied (PL-RM, circles). We also show the performance of a step-index multimode fibre of the same length and mode capacity and the photonic lantern, bent while guide-star based correction is applied (MMF, squares)~\cite{li2021memory}. The power ratio maps of $4$ different points on these curves (i-iv) are highlighted. Here the circles indicate the location of the guide-star on the right hand edge of the core.}
    \label{fig:bend-sense}
\end{figure}

Figure~\ref{fig:bend-sense} shows how the mean power-ratio of distal focussing is reduced as the photonic lantern bends. We find that both bend compensation strategies are highly sensitive to a small degree of bending of the photonic lantern itself. For simulations of our $N_{\text{c}}=98$ mode photonic lantern, we observe that the mean power-ratio decreases substantially for bend radii between $5$\,m - $1$\,m. We see that guide-star assisted bend compensation is marginally more robust to bending of the photonic lantern than reflection-based bend compensation. This is likely due to the fact that the guide-star method relies on well-defined structured light passing only in the forward direction through the photonic lantern, while the reflection-based bend compensation methods rely on well-defined structured light fields passing in both directions.

To put this bend sensitivity in context, we also compare bending of the photonic lantern transition to bending of an equivalent segment of step-index multimode fibre. We simulate the bend sensitivity of fibre-eigenmode based guide-star assisted TM correction (using the method presented in ref.~\cite{li2021memory}) through an ideal $4$\,cm length of multimode fibre of equal $\text{NA}$ and core radius to the output port of our photonic lantern -- see Methods \S7. Our model indicates that, upon bending, the principal lantern modes of the photonic lantern transition decorrelates much more readily than the eigenmodes of the equivalent segment of multimode fibre -- rendering the photonic lantern far more sensitive to bending.

We qualitatively explain this phenomenon as follows. All spatial modes propagating along a multimode fibre must totally internally reflect from the core-cladding boundary, and so span across the entire core region. Thus all multimode fibre eigenmodes simultaneously experience the optical path length changes that occur across the core region when the multimode fibre bends. Meanwhile, MCFs are inherently much more sensitive to bending than multimode fibres, since light propagates within isolated MCF cores that can experience different changes in path length during bending. Photonic lanterns inherit this high level of bend sensitivity, since light entering localised to a single core evanescently couples to surrounding cores within the photonic lantern transition. This process is highly sensitive to the local changes in optical path length that occur when the photonic lantern bends, leading to radically different speckle patterns developing.

In summary, our simulations indicate that the photonic lantern transition itself should be rigidly held for optimal control of distal optical fields when using the MCF bend compensation methods introduced here. This could be achieved by encasing the photonic lantern within a rigid needle, which is compatible with many use cases of this technology. The rigidity of the photonic lantern could also be increased by reducing its length, although this would come at the expense of higher levels of loss. We also note that, due to its short length, there are only a limited number of ways the photonic lantern transition could bend. If this space of possible photonic lantern TMs could be fully mapped out in advance, it may be possible to account for photonic lantern bending by combining our MCF correction concept with strategies that adaptively select from a library of pre-calibrated TMs~\cite{farahi2013dynamic,wen2023single,wen2025cascaded,mekhail2026portable}.\\


\noindent{\bf Discussion}

\noindent We now put our work in the context of other fibre imaging approaches. We recently demonstrated computational imaging through flexible photonic lantern fibres by projecting light through one core at a time to sequentially illuminate the scene with a series of pre-calibrated speckle patterns~\cite{choudhury2020computational,shin2016compressive}. This method has the advantage of being insensitive to bending of the MCF while requiring no active wavefront shaping or bend calibration. However, as shown in Supplementary Information \S2, in this case image reconstruction requires solving a poorly conditioned inverse problem, and so the SNR scales badly with increasing image resolution. Here we have demonstrated high contrast diffraction-limited scanning imaging, which is optimal in terms of SNR, with the potential to boost the signal-to-noise ratio of imaging by orders of magnitude compared to our earlier work.

We next consider guide-star assisted fibre bend correction, which we earlier proposed for multimode fibres~\cite{li2021memory}. There are some subtle differences in how this method performs when applied to photonic lantern fibres. In the multimode fibre case, the optimal guide-star location is at the edge of the core (to best overlap with all fibre eigenmodes), and precise identification of the eigenbasis of the multimode fibre TM is necessary. This eigenbasis is challenging to identify with high fidelity, since unknown imperfections due to the fibre fabrication process must be accounted for, and effects solely due to the multimode fibre itself must be decoupled from the effects of aberrations in the surrounding optical system~\cite{ploschner2015seeing}. Furthermore, when a step-index multimode fibre bends, global phase shifts of the eigenmodes (which can be tracked by a distal guide-star) are accompanied by small changes to the structure of the eigenmodes themselves~\cite{cao2023controlling} (which cannot be tracked by a distal guide-star, as shown in Fig.~\ref{fig:bend-sense}). This means guide-star assisted imaging through multimode fibres may suffer from a reduced field-of-view, an effect likely to be exacerbated for longer fibres. 

Photonic lantern fibres circumvent many of these difficulties, since the bend operator is already diagonal when expressed in the MCF core input basis. Our guide-star or reflection-based bend compensation strategies enable diffraction-limited imaging across the entire distal facet, irrespective of MCF length. A guide-star can be placed anywhere on the distal facet, as every location is equally likely to overlap with the principle lantern modes. Three conditions must be satisfied for our methods to work well: (i) MCF bending must not be extreme enough to significantly couple light between the MCF cores, (ii) the bend operator must remain constant during the time taken to measure it, and (iii) the photonic lantern itself must be rigidly fixed and remain unbent.

Our simulations also reveal a phenomenological difference between guide-star assisted imaging through multimode fibres and photonic lantern fibres: In the multimode fibre case, as mentioned above, unknown changes in the shape of each eigenmode decrease the distal focussing fidelity, and the area where focussing can be achieved collapses into a local region around the guide-star, known as an {\it isoplanatic patch}, as shown in Fig.~\ref{fig:bend-sense} (panel ii)~\cite{li2021memory}. In the case of a photonic lantern fibre, bending of the photonic lantern transition equivalently results in unknown changes to the shapes of the speckle modes, also leading to a loss of distal focussing fidelity. Yet in this case there is no formation of an isoplanatic patch. Instead, the power-ratio decreases approximately uniformly across the whole distal facet, as can be seen in Fig.~\ref{fig:bend-sense} (panel iii).

The root of this difference is in the way the modes couple to one another when they are perturbed. Multimode fibre eigenmodes preferentially couple to other eigenmodes that carry similar levels of orbital angular momentum and radial momentum~\cite{li2021compressively,gutierrez2024characterization,gokay2026explaining}. This preferential coupling leads to the optical field in regions of the field-of-view close to the guide-star being well approximated by the original eigenmodes, while regions further away are less well approximated, i.e., the formation of an isoplanatic patch around the guide-star. Meanwhile, in the case of the photonic lantern fibre, modes of the speckle basis are equally likely to couple into all other modes, so no region of the distal facet is preferentially corrected, irrespective of its proximity to the guide-star. This observation highlights that isoplanatic patch formation is not universal, but is reliant upon preferential mode coupling exhibited by a modal basis when it is randomly perturbed.

Finally, we consider our work in the light of previous reflection based methods that have been employed to compensate for MCF bending. Such techniques typically require a way to couple light from one outbound core to many return cores. Previously proposed methods rely on partial reflectors located at a stand-off distance beyond the distal MCF facet, with the diffraction of light through free-space coupling the light transmitted through a single core in the outbound direction to multiple cores on the return path~\cite{kuschmierz2018self,badt2022real,choi2022flexible}. These approaches have already been scaled to enable coherent reflection imaging through MCFs supporting $\sim$\,$10,000$ cores which can be contorted at video rates.

Here we have introduced a photonic lantern as novel distal core-coupling element in which the optical field is randomly mixed within the fibre itself. By holding the number of spatial modes roughly constant throughout this optical system, we ensure close to full control over the distal output field. Wavefront shaping is compatible with reflection or fluorescence imaging, and we have sketched a path forward to achieve single-shot active MCF bend compensation. A future challenge will be to scale up the mode capacity of our polarisation-maintaining photonic lantern fibre beyond the $98$ cores we demonstrate here. Complementing our work, we also highlight a recent study that explored how weakly tapering the distal facet of an MCF can be used to boost holographic focussing efficiency~\cite{genchi2025two}.
These approaches will further benefit from the emergence of efficient high speed SLMs for fast wavefront control~\cite{tzang2019wavefront,rocha2024fast}.\\


\noindent{\bf Conclusions}

\noindent In summary, we have demonstrated that polarisation-maintaining multicore photonic lantern fibres offer a new route towards overcoming the long-standing challenge of delivering high-fidelity light control and imaging through dynamically flexible optical fibres. The minimal mode-coupling properties of a polarisation-maintaining MCF, combined with the strong mode coupling properties of the photonic lantern transition itself, enable close to full control over the optical field projected to the distal facet, and also allow the bend-induced phases in the MCF to be deduced from information gained from the proximal end.

The remote delivery of complex states of light through flexible photonic lantern fibres has promises a range of applications across areas such as bio-imaging~\cite{becerra2026photonic}, coherent imaging in astronomy~\cite{kim2024coherent}, and spatial mode sorting~\cite{billaud2024tilba}. For example, further development of the techniques we have presented here could lead to imaging within freely mobile animals through metre-scale lengths of MCF via rapid adaptive fibre bend compensation. In the future it will be interesting to explore the control of broadband (few nm) ultra-fast pulses through photonic lantern fibres, potentially enabling non-linear bio-imaging.

Beyond micro-endoscopy, a photonic lantern can also be used to decompose light collected by telescopes into a set of single modes, which opens new opportunities for coherent imaging~\cite{kim2024coherent,kim2025sky,dobias2026wavelength,eikenberry2026decoding}. We anticipate such techniques would benefit from the enhanced polarisation control offered by polarisation-maintaining MCFs photonic lanterns. Moreover, the phase compensation and metrology techniques we have demonstrated could enable stable MCF links from a telescope focal plane to a beam combining instrument. Finally, polarisation-maintaining photonic lanterns could provide a way to efficiently couple free-space multimode light into the set of single mode waveguides for ultrafast processing within photonic integrated circuits and vice versa ~\cite{bogaerts2020programmable,seyedinnavadeh2024determining,billaud2024tilba,lavery2024re}, with applications in areas including free-space optical communications and LiDAR.

\section{Methods}

\noindent{\bf \S1: Transmission matrix measurement}

\noindent The optical system to measure the TM of our photonic lantern fibre is based on a Mach–Zehnder interferometer (see Fig.~\ref{fig:TMBending}(a) for simplified schematic, and Supplementary Fig.~S1(a) for a full diagram of the experimental set up). Laser light of wavelength ${\lambda = 633\,\text{nm}}$ is split into two paths, which we refer to as the `signal arm' and the `external reference arm'. Light in the signal arm first reflects from an SLM before propagating through the photonic lantern fibre. In our case the SLM is a digital micro-mirror device (DMD), which is used as a reconfigurable binary amplitude diffraction grating~\cite{lee1979binary,turtaev2017comparison} to spatially shape both the amplitude and phase of the optical field incident on the MCF facet at the input of the photonic lantern fibre.

The DMD can be programmed to excite individual or combinations of MCF cores with mode-matched beams of independently tunable intensity and relative phase. Transmitted light exiting at the multimode facet of the photonic lantern fibre is imaged onto a camera where it interferes with light from the external reference arm, allowing output fields to be measured using phase-stepping holography.

To measure the TM, $\VecT$, in the principal lantern mode basis, each MCF core is illuminated in turn, with $p=4$ global phase steps of $0,\,\pi/2,\,\pi,\, 3\pi/2$\,rads being sequentially applied to the light illuminating each core. The interference patterns of the transmitted speckles with the reference beam are recorded by the camera and processed to obtain the phase of the speckle patterns. A fifth measurement, with the reference beam blocked, measures the amplitude of each speckle pattern. From these measurements the resulting complex speckle fields generated at the multimode facet are recorded in the pixel basis of a camera imaging the output facet. This data is used to construct $\VecT$ column by column: the $n^{\text{th}}$ output field, generated by illuminating the $n^{\text{th}}$ MCF core, is reshaped into a column vector which forms the $n^{\text{th}}$ column of $\VecT$~\cite{li2021compressively}.

The DMD can be switched at up to $20\,\text{kHz}$, however our present experiment is not optimised for measurement speed, and runs at a binary hologram modulation rate of ${f=50\,\text{Hz}}$, limited by the camera. This leads to a present TM measurement time of ${\tau = (p+1)N_\text{c}/f \sim 10\,\text{s}}$ (excluding the time taken for hologram upload and data processing).

To test the TM, we use it within Eqn.~\ref{eqn:TM_invert} to generate and raster scan a focussed spot across the distal facet, and measure the mean power-ratio. In order to accurately capture the power-ratio of each spot with out 8-bit camera, we perform high dynamic range imaging, by recording the intensity of output fields at a range of exposure times, and combining these images to synthesis images of higher bit-depth. These images enables more reliably calculation of the power in the background and within each focussed spot.\\

\noindent{\bf \S2: Characterising the NA of the photonic lantern fibre}

\noindent The data collected to measure the TM can also be used to estimate the NA of the multimode output port of the photonic lantern fibre: we Fourier transform each measured complex output speckle pattern and find the minimum radius of the far-field disk that encompasses all of these principal lantern modes. In $k$-space this disk has a radius of $k_\text{r} =(k_\text{x}^2 + k_\text{y}^2)^{1/2}$, which relates to the estimated NA via $\text{NA} = k_\text{r}/k$, where $k=2\pi/\lambda$.\\

\noindent{\bf \S3: Transmission scanning imaging and axial refocussing}

\noindent The example transmission images shown in Fig.~\ref{fig:WSPhotonicLantern}(c-e) are recorded as follows. We first measure the TM of the photonic lantern fibre as described above, and raster a focussed spot across the output facet. For each spot location, the required input field is calculated using Eqn.~\ref{eqn:TM_invert}, and these fields are then transmitted sequentially through the fibre. The object at the output of the fibre consists of a mirrored calibration slide that reflects light everywhere except within certain features (e.g., within lines, numbers etc.), where the light is transmitted through the calibration slide. For the spot in each location, we record the total intensity arriving at a camera (i.e, using the camera as a bucket detector with no spatial resolution) which is positioned after the calibration slide, imaging the output end of the photonic lantern fibre -- see Supplementary Fig.~S1(b). To reconstruct an image, we create a map of the measured intensity at the camera as a function of the projected spot location. When the plane of the calibration slide is not the same as the plane in which the foci are raster scanned, the reconstructed image appears out of focus, as each focussed spot defocusses before being partially transmitted/reflected by the slide. This is the case for Fig.~\ref{fig:WSPhotonicLantern}(c).

To axially reposition the plane in which the distal foci are raster scanned across, by a distance $\delta z$, we include in our model of the optical system an additional operator $\VecP\left(\delta z\right)$ that propagates the field a distance of $\delta z$ through free-space at the output of the fibre, such that the output field $\Vecv$ is related to the input field $\Vecu$ by
\begin{equation}\label{eqn:refocus_for}
    \Vecv = \VecP\left(\delta z\right)\cdot\VecT\cdot\Vecu,
\end{equation}
where $\VecP$ is constructed using the angular spectrum method
\begin{equation}
\VecP\left(\delta z\right) = \mathcal{F}^{-1}\cdot\VecD\left(\delta z\right)\cdot\mathcal{F},
\end{equation}
where $\mathcal{F}$ is the Fourier transform matrix and ${ \VecD\left(\delta z\right)}$ is a unitary diagonal matrix describing the phase delay picked up by each plane-wave component of the field upon axial propagation of a distance $\delta z$.   Thus, to calculate the input field $\Vecu_{\text{t}}$ required to focus to the $m^{\text{th}}$ pixel of an axially shifted output plane we use 
\begin{equation}\label{eqn:refocus_back}
    \Vecu_{\text{t}} = \left(\VecP\left(\delta z\right)\cdot\VecT\right)^\dagger\cdot\Vecv_{\text{t}},
\end{equation}
where $\Vecv_{\text{t}}$ is a one-hot vector in which element $m$ is set to 1. Since $\VecP(\delta z=0)$ is equal to the identity matrix, Eqn.~\ref{eqn:refocus_back} collapses to Eqn.~\ref{eqn:TM_invert} for no axial shift of the image plane.

Here we have demonstrated transmission imaging due to the low power of the laser used in our experiments ($\sim1\,\text{mW}$), however we note that the image reconstruction method does not rely on the spatial resolution of the camera and works in the same way as reflection or fluorescence imaging through the fibre, where instead of measuring the total transmitted intensity at the output, the total reflected intensity is measured back at the input of the fibre, as demonstrated, for example, in refs.~\cite{turtaev2018high,ohayon2018minimally,leite2021observing,stibuurek2023110}.\\

\noindent{\bf \S4: Remotely focussing on a guide-star}

\noindent Here we describe in more detail how the input field required to focus on a tip-mounted guide-star can be determined with access only to the input end of the fibre. A fluorescent particle's emission is linearly proportional to the excitation intensity. Therefore, measuring the integrated fluorescent emission, back at the input end of the fibre, emitted by a tip-mounted fluorescent guide-star, reports the local excitation intensity on the guide-star's location.

To find the input field that focusses onto the guide-star, $\Vecu_\text{gs}$, the process is similar to TM measurement: the photonic lantern fibre is probed in the MCF core basis by exciting each core sequentially in a pair-wise fashion with a common reference core, which provides a common-path reference beam. We refer to the two illuminated MCF cores as the `test' core, and the `reference' core. Phase shifting holography is used to recover the global phase of each corresponding output field on the location of the guide-star. Light in the reference core co-propagates through the photonic lantern alongside light from test core $n$, and the guide-star is illuminated by the resulting interference pattern formed by the two speckle modes at the output. Cycling the global phase of the reference mode through $2\pi$\,rad, with respect to the test mode $n$, causes a sinusoidal change in excitation intensity on the guide-star, and a corresponding sinusoidal variation in the fluorescent intensity signal measured back at the proximal end of the fibre. The phase of this sinusoid reveals the relative phase between the reference mode and the test core mode $n$, on the position of the guide-star. Repeating this measurement for all input modes, with the same reference core, reveals the phase of all elements of the $m_0^{\text{th}}$ row of the TM of the fibre in the new state, $\VecT'$, where $m_0$ indexes the lateral position of the guide-star on the output facet~\cite{vcivzmar2010situ}. This information is then used to calculate the bend operator of the fibre in its new state, as described in the main text.

In our proof-of-principle experiment, to avoid the additional complexity of engineering a guide-star on the end of the fibre facet, we instead demonstrate the principle using the signal recorded from a single camera pixel, where the camera images the output facet of the fibre. Supplementary Fig.~S2(a) shows a schematic of the experimental set-up used in this case. This emulation requires illumination of the photonic lantern fibre with the same set of input modes, and follows the same analysis of the measured signals, as would be carried out using a real fluorescent guide-star with proximal fluorescence detection.\\

\noindent{\bf \S5: Common-path referenced reflection measurements}

\noindent  Our common-path referenced reflection-based bend compensation method relies on an identical sequence of input beams to that used for guide-star assisted bend compensation, while the detection scheme and analysis are different. The full optical setup used in this experiment is shown in Supplementary Fig.~S2(b). Part of the reflection matrix of the photonic lantern fibre is probed in the MCF core basis by exciting each core sequentially in a pair-wise fashion with a common reference core. The detection scheme is based on measurements of the intensity of light reflected from the distal multimode facet back to the input: the phase of the light propagating down the reference core is stepped through $2\pi$ rad, with respect to light propagating down the test core, and the sinusoidal variation in the intensity of light emanating back out of a third `detection' core is monitored using a camera imaging the input MCF facet. The phase of this sinusoid reveals the phase of the test core light. In this way, exciting the $m^{\text{th}}$ test MCF core and observing the light reflected back through the $n^{\text{th}}$ detection MCF core enables measurement of $\arg(R^0_{n,m})$ or $\arg(R'_{n,m})$.

Intuitively, we can understand this process as follows: for each measurement, light propagates outbound through two separate cores (the reference core and the core $m$ under test) -- thus experiencing core-dependent phase shifts. The photonic lantern transition mixes the light from both outbound paths, and a portion of this light returns along the same core (the detection core $n$) -- thus ensuring the measurement is insensitive to core-dependent phase shifts on the return journey. Therefore, the relative phase-shift picked up by light in different cores solely in the forward direction can be measured. Equation~\ref{eq:reflPhiRow} calculates the difference between these core-dependent phase changes with the MCF in the new and original configurations -- revealing the phase of the diagonal elements of the bend operator. We find the core-dependent phase shifts up to an unknown global phase factor of $\delta\phi_n$ which arises from the unknown change in phase of light propagating back along the detection core when the MCF bends. In practice, there is also an unknown global phase change applied to all measurements, due to the unknown phase of the light transmitted through the reference core to the distal facet. $\delta\phi_n$, and the unknown phase change of the light in the reference core, can be set to zero when applying Eqn.~\ref{eq:reflPhiRow} since they take the same values for all core measurements.\\

In principle, it is only necessary to observe light reflected back along a single detection MCF core, as described above. However for each measurement it is possible to simultaneously monitor light reflected back through all cores using the proximal camera. Therefore, to enhance the SNR, we can average over different measurements of the bend operator phases, where each measurement is based on different detection cores. Thus, the measurement-averaged phase of the $m^{\text{th}}$ element of the bend operator, $\overline{\delta\phi}_m$, is given by:
\begin{equation}\label{eq:reflIntAv}
    \overline{\delta\phi}_m = \arg\left[\sum_n\left|R^0_{n,m}\right|\text{e}^{i\delta\phi_m^n}\right],
\end{equation}
where $\delta\phi_m^n$ is the phase obtained when testing core $m$ and detecting light reflected along core $n$, i.e., ${\delta\phi_m^n = \arg(R'_{n,m}) - \arg(R^0_{n,m})}$. This is the approach we took in our experiments.\\

\noindent{\bf \S6: Externally referenced reflection measurements}

\noindent The full optical setup for our externally referenced bend compensation experiment is shown in Supplementary Fig.~S2(c). Intuitively, we can understand this bend correction process as follows: irrespective of the bend state of the MCF, illuminating a single core always generates the same core-dependent complex speckle pattern at the multimode waveguide output (except for an unknown bend-dependent global phase shift). Upon reflection, this speckle pattern serves as a well-defined mode launched from the distal end of the fibre to probe the current bend state of the MCF, allowing the relative phase changes of light propagating along each core $n$ to be directly measured at the proximal end of the fibre.

Although this bend correction measurement can, in principle, be made for all cores in a single shot, to improve the SNR we averaged over the phases measured when several different MCF cores were sequentially illuminated (indexed by $m)$. Here the measurement-averaged phase of the $n^{\text{th}}$ element of the bend operator, $\overline{\delta\phi}_n$, is given by:
\begin{equation}\label{eq:reflIntAv2}
    \overline{\delta\phi}_n = \arg\left[\sum_m\left|R^0_{n,m}\right|\text{e}^{i\delta\phi_n^m}\right],
\end{equation}
where $\delta\phi_n^m$ is the phase obtained when illuminating core $m$ and detecting light reflected back along core $n$, i.e., ${\delta\phi_n^m = \arg(R'_{n,m}) - \arg(R^0_{n,m})}$. This approach comes at the cost of increasing the number of sequential measurements that need to be made -- but this number can be far lower than the the number of MCF cores ($5$ cores were sequentially illuminated in our case).\\


\noindent{\bf \S7: Photonic lantern bend modelling}

\noindent We model the photonic lantern transition using in-house custom written software based on the finite-difference beam propagation method (FD-BPM)~\cite{pedrola2015beam} with an alternating-direction implicit (ADI) scheme and perfectly matched layer (PML) absorbing boundary conditions~\cite{huang1996perfectly}. Here we use the scalar approximation, which is well suited to our polarisation-maintaining photonic lantern fibre. We first simulate the TM of a straight photonic lantern equivalent to that used in our experiment: transforming an MCF with ${N_\text{c}=98}$ single mode cores into a multimode waveguide over a length of 4\,cm. This model yields the forward TM of the photonic lantern, ${\VecT(r=\infty)}$, where here $r$ is the radius of curvature through which the photonic lantern is bent (i.e.\ an infinite radius of curvature in the perfectly straight case). The columns of ${\VecT(r=\infty)}$ represent the transmitted speckle pattern associated with the illumination of each MCF core. We next simulate the TM of the photonic lantern for a range of different bend radii, the set values given by ${r = \{100,10,5,1,0.05,0.01\}\,\text{m}}$, yielding a set of TMs as a function of bend radius, ${\VecT(r)}$. 

We now apply our guide-star and reflection-based bend compensation strategies to our bent photonic lantern datasets. These strategies operate under the assumption that bending only induces global phase changes to the transmitted speckle patterns of the speckle basis. Yet bending of the photonic lantern itself also changes the structure of the transmitted speckle patterns, as well as their global phase. Therefore we study how the fidelity of light control at the output of the photonic lantern is reduced as the assumption in our bend compensation strategies breaks down when the photonic lantern itself bends.

Each bend compensation method assumes the TM of the system is given by Eqn.~\ref{eq:bendOp}, yielding an approximation to the true TM of the photonic lantern under bending: ${\VecT_{\text{est}}(r) = \VecT(r=\infty)\cdot\VecB(r)}$. Here the bend operator is unitary and diagonal as before, meaning it can only represent global phase changes to the initial speckle basis upon bending of the photonic lantern (which will depend upon bend radius $r$), but $\VecB$ cannot capture any mode coupling that result in changes to the structure of the distal speckle patterns. We assess the validity of this approximation by calculating ${\VecT_{\text{est}}(r)}$, using it to find the input field to attempt to create a focus at the multimode output facet, and then propagating this field through the real TM of the bent photonic lantern, ${\VecT(r)}$. For example, when attempting to generate a target focus ${\Vecv_{\text{target}}}$ at the output of a photonic lantern with a bend radius of $r$, we actually generate ${\Vecv_{\text{actual}}}$ given by:
\begin{equation}\label{eq:lantBend}
    \Vecv_{\text{actual}}=\VecT(r)\cdot\VecT^\dagger_{\text{est}}(r)\cdot\Vecv_{\text{target}}.
\end{equation}
To characterise the effects of bending of the photonic lantern as a function of bend radius $r$ and bend compensation strategy, we use Eqn.~\ref{eq:lantBend} to simulate attempting to raster scan a foci across the output facet, and calculate the mean power ratio, which is shown in Fig.~\ref{fig:bend-sense}.

For a comparison with guide-star assisted imaging through multimode fibres shown in Fig.~\ref{fig:bend-sense}, we model the bending of an ideal multimode fibre using the methods described in ref.~\cite{ploschner2015seeing}, and apply guide-star assisted TM correction as described in ref.~\cite{li2021memory}, assuming perfect knowledge of the ideal multimode fibre eigenmodes when the fibre is straight.

\bibliography{bendCompRefs}

\section{Acknowledgements}

DBP acknowledges financial support from the European
Research Council (ERC): ERC Starting grant {\it PhotUntangle}, no. 804626; and ERC Consolidator grant {\it ModeMixer}, no. 101170907. DBP also thanks the UK Engineering and Physical Sciences Research Council (EPSRC) for financial support (EP/Z535928/1). RJK was supported by a Doctoral Training Partnership grant from the EPSRC (EP/T518049/1). RRT acknowledges support from the UK Science and Technology Facilities Council (ST/V000403/1), and the EPSRC (EP/X03299X/1, EP/S000410/1). AM acknowledges support via an EPSRC funded PhD studentship. RRT, TAB and KH thank the UK EPSRC for funding via the ``U-care" project, EP/T020903/1. TAB and TW acknowledge funding from the European Commission Horizon 2020 project OPTICON (730890). JMS thanks the EPSRC for supporting the development of the polarisation-maintaining multi-core fibre (EP/K020102/1). RJK and DBP thank Shuhui Li for early work on the experimental setup and software. JCAR and DBP thank Joel Carpenter for support. RRT thanks Jonathan Leach for useful discussions.

\section{Contributions}

RRT and TAB initiated the project. JMS fabricated the polarisation-maintaining multi-core fibre. KH and TAW fabricated the photonic lanterns with support from TAB. DBP proposed the guide-star method. RRT proposed the common-path reference based reflection method. RRT and DBP devised the external reference based reflection method. All presented experimental results were recorded at the University of Exeter, and further experimental investigations were undertaken at Heriot-Watt University. RJK built all variants of the Exeter experimental setup, developed the light control software, and conducted all experiments, with support from CKM and DBP. AM built the Heriot-Watt experimental setup which was used to perform preliminary common-path reflection measurements, with input from AB, HP and RRT. TAW, KH and AM visited Exeter to support experiments. RJK and AM processed experimental data to extract bend operators: guide-star analysis was carried out by RJK, common-path referenced reflection analysis was led by AM, and externally referenced reflection analysis was conducted by RJK and AM. RJK experimentally tested the performance of the bend operators and conducted power-ratio measurement and analysis. JCAR developed the custom FD-BPM code and conducted simulations shown in Fig.~5, and prepared the photonic lantern rendering in Blender for Fig.~1. DBP and RJK developed the mathematical description of the system, created the figures, and wrote the paper, with text also provided by RRT and AM. All other authors provided editorial input.

\section{Disclosures}

The authors declare the following competing interests: TAB and RRT are named inventors on granted patents and pending patent applications (JP7249647B2, US11696676B2, KR102588057B1, CN110831478B, EP3599974A1) relating to some of the technologies and methods described in this paper. The remaining authors declare no competing interests.

\onecolumngrid

\clearpage

\begin{center} {\large {\bf Adaptive wavefront shaping through flexible multicore photonic lantern fibres\\ -- Supplementary Information}} \end{center}

\vspace{1cm}

\twocolumngrid

\setcounter{figure}{0}
\renewcommand{\thefigure}{S\arabic{figure}}

\setcounter{equation}{0}
\renewcommand{\theequation}{S\arabic{equation}}

\begin{figure*}[t]
    \centering
    \includegraphics[width=0.8\linewidth]{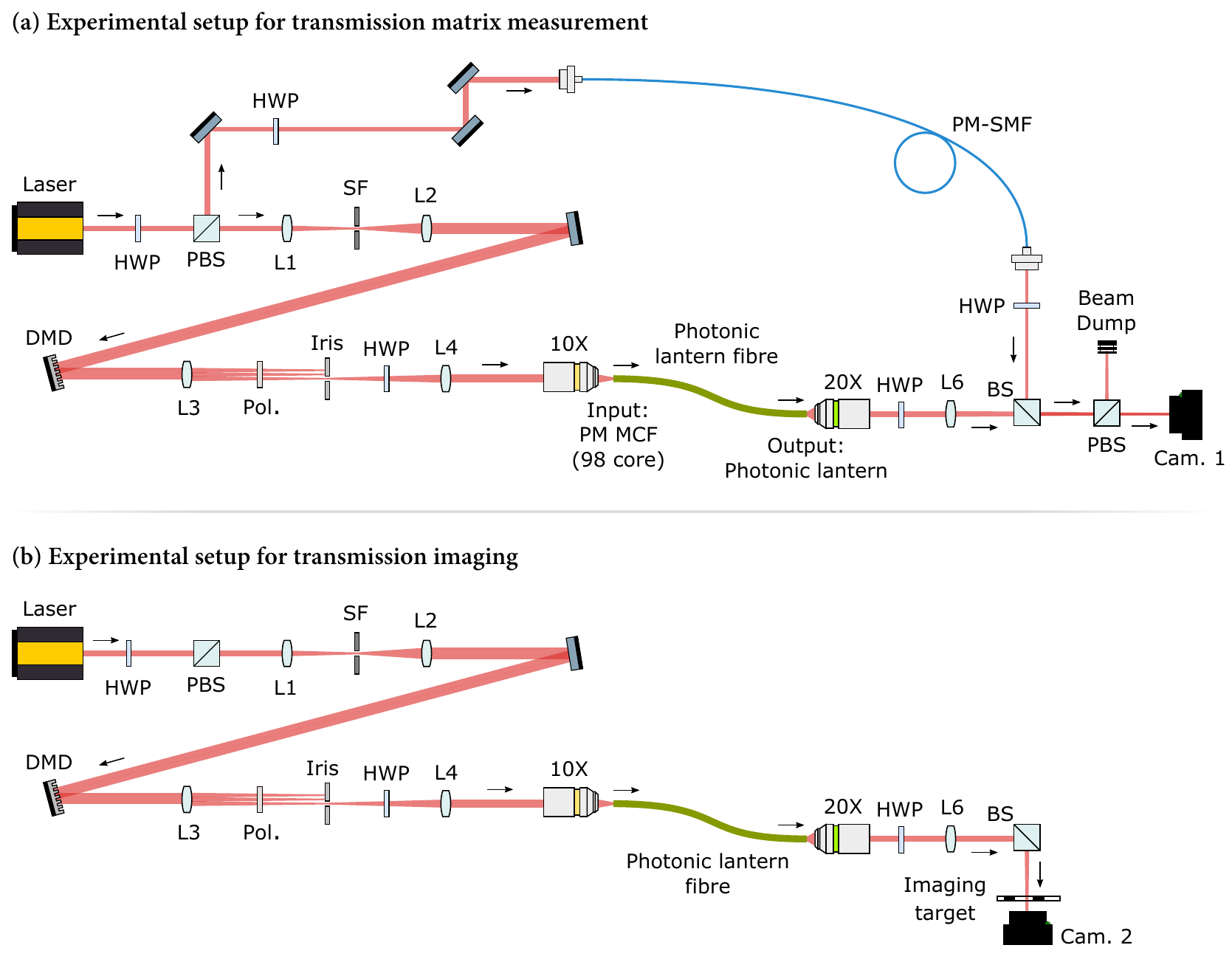}
    \caption{{\bf Experimental setups for TM characterisation and transmission imaging}. (a) TM measurement. (b) transmission imaging. ${\text{HWP = half-wave plate}}$, ${\text{PBS = polarising beam-splitter}}$, ${\text{SF = spatial filter}}$ (diameter of ${50\,\mu\text{m}}$), ${\text{Pol.\ = linear polariser}}$, ${\text{PM-SMF = polarisation maintaining single mode fibre}}$. The focal length of the lenses are: L1 = $50$\,mm, L2 = $400$\,mm, L3 = $300$\,mm, L4 = $200$\,mm, L5 = $200$\,mm, L6 = $200$\,mm (note: lens L5 is only shown in Figs~\ref{fig:BendCompSetup}(b,c)). The laser is a linearly polarised 0.8\,mW HeNe laser of wavelength 633\,nm  (Thorlabs HNLS008L-EC). The DMD has a display of $1024\times768$ pixels and a pixel pitch of $13.7\,\mu\text{m}$, with a maximum switching rate of 22.727\,kHz (Vialux V-7001 VIS). High-speed cameras Cam.\ 0, Cam.\ 1 and Cam.\ 2 are all of maximum resolution $640\times480$ pixels (Basler piA640-210gm).}
    \label{fig:TMSetup}
\end{figure*}

\noindent{\bf \S1: Experimental setups}\\
Figure~\ref{fig:TMSetup}(a) shows a schematic of the experimental setup to measure the TM of the photonic lantern fibre. The underlying setup is similar to that used in our previous work~\cite{li2021memory}. A $\sim1\text{mW}$ linearly polarised HeNe laser beam of wavelength $633\,\text{nm}$ is split into a reference arm (upper beam path) and signal arm (lower beam path). The relative power sent along the two beam paths can be adjusted by rotating a half-wave plate which changes the linear polarisation incident on a polarising beam splitter. The linearly polarised beam in the signal arm is spatially filtered and expanded (using a 4f system consisting of lenses L1 and L2) to fill the display of a DMD.

The binary pattern displayed on the DMD is designed to encode the target beam shape within the first diffraction order -- enabling a single core or multiple cores of the MCF to be illuminated with the desired relative amplitude and phase. Lens L3 Fourier transforms the beam from the DMD to an iris, which selects only this first diffraction order and blocks all other orders. The beam diffracted from the DMD also passes through a polariser to ensure transmitted light remains linearly polarised (i.e., any light scattered into the orthogonal polarisation by diffraction from the edges of the DMD mirrors is filtered out). The shaped beam at the plane of the iris is re-imaged and de-magnified, via a 4f system consisting of lens L4 and a $10\times$ objective lens, to the multi-core input facet of the photonic lantern fibre. The orientation of the linear polarisation is rotated to match the principle axis of the polarisation maintaining MCF cores via a half-wave plate.

The beam emanating from the multimode output of the photonic lantern fibre is magnified and re-imaged onto a high-speed camera (Cam.\ 1), via a 4f system consisting of a $20\times$ objective lens and lens L6. The beam in the reference arm is transmitted through a polarisation-maintaining single mode fibre and recombined with the signal arm via a beam splitter.


\noindent Half-wave plates in both the signal and reference arm, and a polarising beam splitter allow the relative power between the two beams to be tuned. The combined signal and reference beams interfere on the camera and enable holographic field reconstruction and TM measurement, which is described in detail in Methods \S1.

Figure~\ref{fig:TMSetup}(b) shows a schematic of the experimental setup used for transmission imaging. The setup is the same as that used for TM measurement with a small modification. Here the beam transmitted through the photonic lantern fibre is projected onto an imaging target (in this case a calibration sample) via the 2nd beam-splitter output port. The total transmitted intensity for each projected beam is recorded by a high-speed camera (Cam.\ 2). Methods \S3 describes the transmission imaging protocol in detail.

\begin{figure*}[t]
    \centering
    \includegraphics[width=0.8\linewidth]{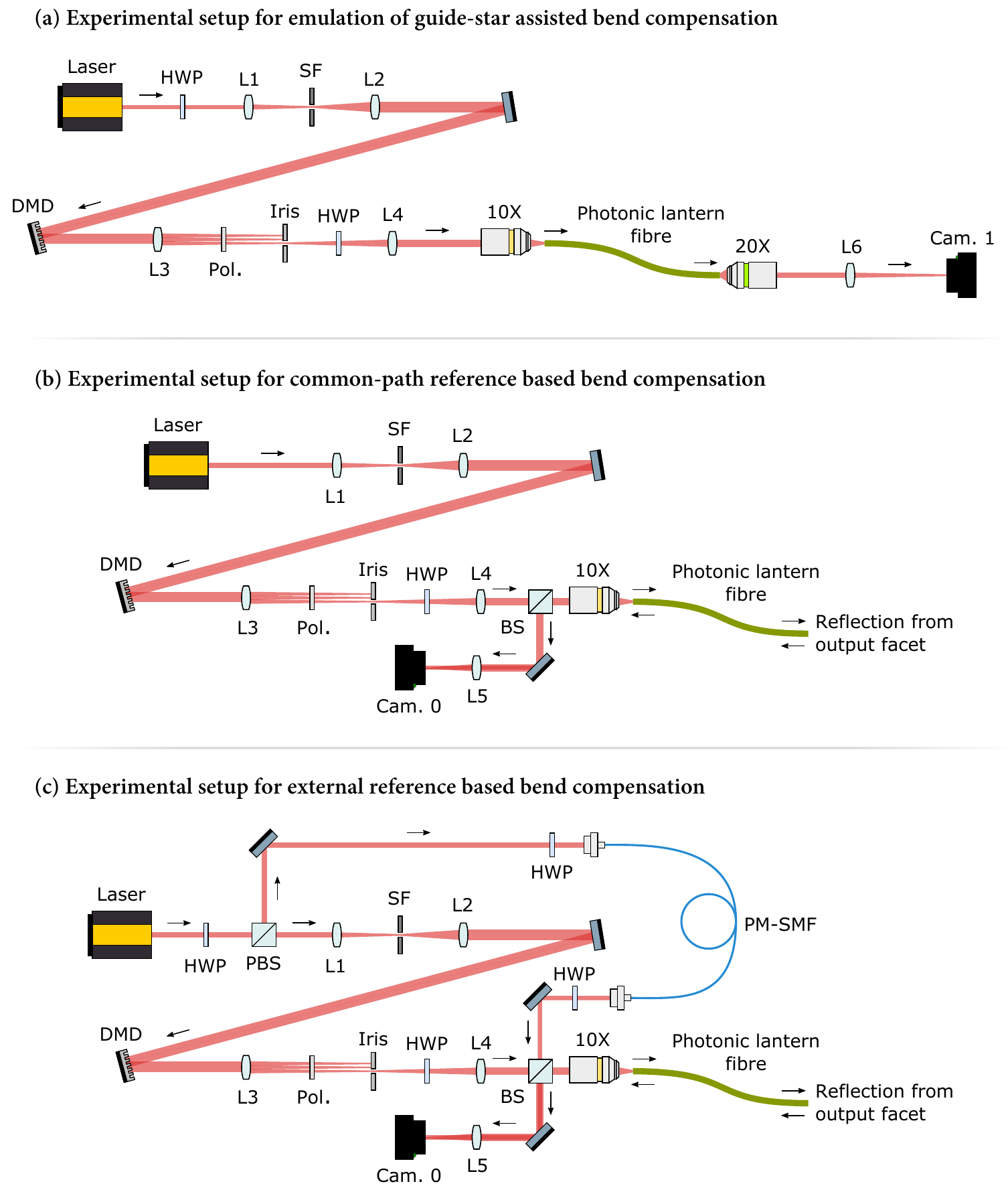}
    \caption{{\bf Experimental setups for bend compensation}. (a) Emulation of guide-star assisted bend compensation. (b) Common-path reference-based bend compensation. (c) External reference-based bend compensation.}
    \label{fig:BendCompSetup}
\end{figure*}

Figure~\ref{fig:BendCompSetup}(a) shows the experimental setup to emulate guide-star assisted bead compensation. Here the intensity recorded by a single pixel of the camera (Cam.\ 1) is used to calculate the bend operator as described in Methods \S4.

Figure~\ref{fig:BendCompSetup}(b) shows the experimental setup for common-path reflection-based bead compensation. Light reflects from the distal facet of the photonic lantern fibre. A camera (Cam.\ 0) images the input MCF facet of the photonic lantern fibre, recording reflected light emanating back out of the input facet. Methods \S5 gives detail of the protocol used to calculate the bend operator. 

Figure~\ref{fig:BendCompSetup}(c) shows the experimental setup for externally referenced reflection-based bead compensation. Light reflected from the distal facet of the photonic lantern fibre is combined with light from an additional reference arm to interfere on a camera (Cam.\ 0) allowing the bend operator to be recovered as described in Methods \S6.\\

\noindent{\bf \S2: SNR of point scanning vs.\ computational imaging}

\noindent It is well known that point scanning optimizes the SNR for an imaging system that projects a sequence of illumination fields onto a scene. Here we sketch out a brief explanation of why this is the case. 

Our task is to reconstruct an image of a scene with reflectivity as a function of position given by $\Veco$, which represents the scene regularly sampled at a given resolution with discrete pixels. In the absence of noise, when the $i^{\text{th}}$ patterned light field $\Vecp_i$ illuminates this scene, the total intensity of reflected light, which we record as measurement $m_i$, is given by
\begin{equation}
m_i = \Vecp_i^\top\cdot\Veco,    
\end{equation}
where here $\Veco$ and $\Vecp_i$ are expressed as column vectors -- flattened versions of the 2D arrays representing the scene and the projected pattern, and $^\top$ is the transpose operator. By collecting all projected patterns into the rows of matrix $\VecP$ (such that the $i^{\text{th}}$ row of $\VecP$ is given by $\Vecp_i^\top$), we can write 
\begin{equation}\label{Eqn:imageForward}
    \Vecm = \VecP\cdot\Veco,
\end{equation}
where column vector $\Vecm$ stores the measurements (the $i^{\text{th}}$ element of $\Vecm$ being given by $m_i$). Thus an image of the scene can be reconstructed by multiplying both sides of Eqn.~\ref{Eqn:imageForward}
by the inverse of $\VecP$, yielding 
\begin{equation}
    \Veco = \VecP^{-1}\cdot\Vecm,
\end{equation}
where here we assume that matrix $\VecP$ is full rank. 

A more realistic expression for the imaging model, now including detector noise, is given by
\begin{equation}\label{Eqn:imageForwardNoise}
    \Vecm' = \VecP\cdot\Veco + \Delta\Vecm,
\end{equation}
where column vector ${\Vecm'=\Vecm+\Delta\Vecm}$ are the measurements incorporating unknown additive noise given by column vector $\Delta\Vecm$. Multiplying both sides of Eqn.~\ref{Eqn:imageForwardNoise} by the inverse of $\VecP$ leads to an approximate image ${\Veco'=\Veco+\Delta\Veco}$, which differs from the actual image $\Veco$ by ${\Delta\Veco = \VecP^{-1}\cdot\Delta\Vecm}$:
\begin{equation}
    \Veco' = \VecP^{-1}\cdot\Vecm' = \Veco + \VecP^{-1}\cdot\Delta\Vecm.
\end{equation}

To understand how the choice of projected measurement patterns impact the SNR of the final image, it is instructive to inspect the properties of real matrix $\VecP$ by calculating its singular value decomposition: ${\VecP = \VecU\cdot\VecD\cdot\VecV^{\dagger}}$.
Here real orthogonal matrices $\VecU$ and $\VecV$ express the left and right hand singular vectors on their columns, and diagonal real matrix $\VecD$ holds the singular values along its diagonal. The pseudo-inverse of $\VecP$ is given by ${\VecP^{-1} = \VecV\cdot\VecD^{-1}\cdot\VecU^{\dagger}}$. In particular, the condition number $\kappa$ of matrix $\VecP$ is given by the ratio of the largest to smallest singular values (given by $\sigma_{\text{max}}$ and $\sigma_{\text{min}}$ respectively):
\begin{equation}
    \kappa\left(\VecP\right) = \frac{\sigma_{\text{max}}\left(\VecP\right)}{\sigma_{\text{min}}\left(\VecP\right)}
\end{equation}
The condition number gives the asymptotic worst case change in the output for a relative change in the input, i.e., it is a measure of the sensitivity of $\VecP$ to small changes in the input which may be introduced by measurement noise~\cite{skare2000condition}. By propagation of the errors it can be shown via the Cauchy–Schwarz inequality that the relative error in the reconstructed image is related to the relative error in the measurements via the condition number of matrix $\VecP$:
\begin{equation}
    \frac{\norm{\Delta\Veco}}{\norm{\Veco}}\leq\kappa\left(\VecP\right)\frac{\norm{\Delta\Vecm}}{\norm{\Vecm}},
\end{equation}
where $\norm{\cdot}$ represents the matrix 2-norm.

Therefore, under the assumption that any choice of projected illumination patterns can be achieved with the same total illumination power, the measurement basis should be ideally chosen so as to minimise the condition number of matrix $\VecP$. Restricting ourselves to positive real functions (i.e., that are achievable by projecting patterns of intensity onto the scene), the condition number $\kappa\left(\VecP\right)$ is minimised when ${\VecP=\VecI}$ (the identity matrix), which is equivalent to reconstructing an image by scanning a focussed spot across the sample. In this case, ${\kappa\left(\VecI\right)=1}$ (notably independent of the number of image pixels $P$ -- although the time to scan all pixels will increase with resolution). This basis also has the added benefit of affording a computationally simple reconstruction, requiring no matrix inversion, since ${\VecI^{-1}=\VecI}$.

In general, sparse orthonormal bases tend to lead to lower condition numbers. By comparison, projecting positive binary patterns with roughly equal areas of bright and dark regions (for example derived from the elements of the Hadamard basis~\cite{pratt1969hadamard}) leads to a condition number of $\kappa\left(\VecP\right) = P$, with progressively noisier image reconstructions as the number of pixels $P$ increases. Similarly, projecting sets of known speckle patterns (such as the speckle basis measured in our work by illuminating each MCF core in turn, and used as an intensity pattern projection basis in our previous work~\cite{choudhury2020computational}) also results in a condition number that increases with $P$ and is non-optimal in terms of SNR, due to the lack of sparsity. This motivates the development of a spot-scanning photonic lantern imaging system in our present work.

\end{document}